\documentclass[preprint,prd,aps,a4paper,10pt,titlepage,superscriptaddress,showpacs]{revtex4-1}
\usepackage{ucs}
\usepackage[utf8x]{inputenc}
\usepackage{amsmath}
\usepackage{amsfonts}
\usepackage{amssymb}
\usepackage{amsthm}
\usepackage[english]{babel}
\usepackage[T1]{fontenc}
\usepackage{graphicx}
\usepackage[dvips]{hyperref}

\begin{document}
\title{Study of the discovery potential for hidden photon emission at future electron scattering fixed target experiments}
\author{T. Beranek}
\author{M. Vanderhaeghen}
\affiliation{Institut f\"ur Kernphysik, Johannes Gutenberg-Universit\"at Mainz, D-55099 Mainz}
\affiliation{PRISMA Cluster of Excellence, Johannes Gutenberg-Universit\"at Mainz, D-55099 Mainz}
\date{\today}

\newcommand{\fslash}[1]{(\gamma\cdot{#1})}
\newcommand{\Abs}[1]{|\vec{#1}\,|}
\newcommand{\Absp}[1]{|\vec{#1}^{\,\prime}|}
\newcommand{\vecp}[1]{\vec{#1}^{\,\prime}}
\newcommand{\ubar}{{\overline{u}}}
\newcommand{\vbar}{{\overline{v}}}
\newcommand{\Ap}{{\gamma^\prime}}
\newcommand{\MeV}{\mathrm{MeV}}
\newcommand{\GeV}{\mathrm{GeV}}
\newcommand{\D}{\mathrm{D}}
\newcommand{\X}{\mathrm{X}}

\begin{abstract}
Electron scattering fixed target experiments are a versatile tool to explore various physics phenomena.
Recently these experiments came into focus to search for $U(1)$ extensions of the Standard Model of particle physics at low energies.
These extensions are motivated from anomalies in astrophysical observations as well as from deviations from Standard Model predictions, such as the discrepancy between the experimental and theoretical determination of the anomalous magnetic moment of the muon.
They also arise naturally when the Standard Model is embedded into a more general theory.
In the considered $U(1)$ extensions a new, light messenger particle $\gamma^\prime$, the hidden photon, couples to the hidden sector as well as to the electromagnetic current of the Standard Model by kinetic mixing, which allows for a search for this particle e.g. in the invariant mass distribution of the process $e (A,\,Z)\rightarrow e (A,\,Z) l^+ l^-$.
In this process the hidden photon is emitted by bremsstrahlung and decays into a pair of Standard Model leptons.
In this work we study the applicability of the Weizs\"acker-Williams approximation to calculate the signal cross section of the process, which is widely used to design such experimental setups.
Furthermore, based on a previous work, we investigate the discovery potential of future experimental setups at the Jefferson Lab.
For that purpose we simulate the relevant cross sections for the signal and the QED background in the actual kinematical setups and obtain projected exclusion limits.
\end{abstract}

\maketitle

\section{Introduction}
Extensions of the Standard Model (SM) by an additional abelian $U(1)$ gauge group were established decades ago \cite{Weinberg:1977ma,Okun:1982xi,Holdom:1986eq,Fayet:1990wx}.
Recently such extensions, which arise automatically, when the SM is embedded into a more general theory such as in particular models of string theory or super symmetry \cite{Pospelov:2008zw,Jaeckel:2010ni,Andreas:2011in}, drew attention motivated by anomalies in astrophysical data \cite{Strong:2005zx,Adriani:2008zr,Cholis:2008wq}.
These anomalies can be explained by annihilating dark matter, where the sector of dark matter is linked to the SM sector by an $U(1)$ mediator \cite{ArkaniHamed:2008qn,Pospelov:2008jd}.
In a minimal model the hidden sector, which contains the dark matter, does not interact directly with the SM matter, but by the exchange of a so-called hidden photon $\Ap$.
The interaction between the hidden photon $\Ap$ and the SM particles, which extends the SM gauge group $SU(3)_c \times SU(2)_L \times U(1)_Y$, can be realized by kinetic mixing between the hidden photon gauge group $U^\prime(1)$ and the SM hypercharge gauge group $U(1)_Y$, where heavy particles charged under both gauge groups are exchanged in a loop \cite{Holdom:1986eq}.
At low energies this gives rise to an effective interaction Lagrangian \cite{Holdom:1986eq} of the hidden photon with the electromagnetic SM current
\[
 \mathcal{L}_{\text{int}}=i\,\varepsilon e\, \bar{\psi}_\text{SM}\, \gamma^\mu\, \psi_\text{SM} \,A^\prime_\mu,
\]
where $A^\prime$ denotes the $\Ap$ field and $\varepsilon$ is the kinetic mixing factor parameterizing the coupling strength relative to the electric charge $e$.
The free parameters in this model are the coupling strength $\alpha^\prime = \varepsilon^2 \alpha$ with $\alpha = e^2/(4\pi) \simeq 1/137$ and the mass of the hidden photon $m_\Ap$, which can be generated e.g. by the Higgs mechanism.
The hidden photon mass $m_\Ap$ can be estimated to be in the range of $10\, \MeV$ to a few $\GeV$ \cite{Fayet:2007ua,Cheung:2009qd,Essig:2009nc}.
In addition, the kinetic mixing factor $\varepsilon^2={\alpha^\prime}/{\alpha}$ is predicted from various models to be in the range $10^{-12} < \varepsilon < 10^{-2}$ \cite{Essig:2009nc,Goodsell:2009xc}.
In pioneering works several constraints on these parameters from existing data were obtained, e.g. data of beam dump searches or the BABAR experiment, as well as from $(g-2)$ analyses \cite{Pospelov:2008zw,Bjorken:2009mm}.
Furthermore, in Ref.~\cite{Bjorken:2009mm} a feasibility study to search for hidden photons in low energy electron scattering fixed-target experiments was performed.
It was found, that the predicted coupling of the $\Ap$ to SM particles and its mass range allows for the $\Ap$ search by accelerator experiments at modest energies with high intensities.
Collider experiments are ideally suited for higher $\Ap$ masses, whereas fixed-target experiments with their high luminosities are ideally suited to probe the $\Ap$ hypothesis in the MeV to $1$ GeV range \cite{Bjorken:2009mm,Batell:2009yf,Batell:2009di,Essig:2009nc,Reece:2009un}.
Several experimental programs are currently underway to search for new light, hidden gauge bosons using fixed-target experiments, e.g. the A1 experiment at the Mainz Microtron (MAMI) \cite{Merkel:2011ze,Beranek:2013yqa}, the APEX \cite{Essig:2010xa,Abrahamyan:2011gv}, HPS \cite{HPS} and DarkLight \cite{Freytsis:2009bh,DarkLight,Kahn:2012br} experiments at the Jefferson Lab (JLAB), as well as future experiments at the MESA facility \cite{Aulenbacher:2012tg,Beranek:2013yqa}.
A1 and APEX have already published first results.
In a variety of recent publications, constraints on the $\Ap$ parameter space from the analysis of beam dump searches \cite{Blumlein:2011mv,Gninenko:2011uv,Gninenko:2012eq,Andreas:2012mt}, meson decays and collider experiments \cite{Aubert:2009cp,Archilli:2011zc,Babusci:2012cr,Echenard:2012iq,Gninenko:2013sr} as well as from other arguments were given \cite{Davoudiasl:2012ig,Endo:2012hp}.
In addition, many other experiments were proposed to probe the light hidden sector or are already underway, for a review see e.g. Refs.~\cite{Hewett:2012ns,Essig:2013lka}.

Although the existing and proposed electron scattering fixed-target experiments differ in their particular setup, the physical process investigated in the experiments is always the same.
The electromagnetic process of scattering an electron beam off a fixed target, which is either a proton or a heavy nucleus, induces the bremsstrahlung emission of a hidden photon as signal or a SM photon as background subsequently decaying into a pair of SM leptons.
The decay particles are detected and their invariant mass is reconstructed, which allows one to search for a bump in the invariant mass spectrum caused by the hidden gauge boson.
The $\Ap$ will manifest itself by a very sharp peak, while the radiative background resulting from the corresponding QED process is described by a smooth distribution.
To summarize, the underlying process
\[
 e+(A,Z)\rightarrow e+(A,Z)+(l^+l^-)
\]
is investigated with respect to an excess of events in a single bin of the invariant mass distribution of the lepton pair.
By this method one will either find a signal of the hidden photon or will be able to exclude regions of the $\Ap$ parameter space described by the kinetic mixing factor $\varepsilon$ and its mass $m_{\Ap}$.

In a previous work \cite{Beranek:2013yqa} we have studied the relevant signal and background processes associated with such experiments.
We have found, that our calculations and the data taken at MAMI are in good agreement.
Based on these results we apply the methods of Ref.~\cite{Beranek:2013yqa} to study the reach of the future HPS \cite{HPS} and DarkLight experiments in detail.

This work is structured as follows:
In Sec.~\ref{sec:method} we summarize the method to calculate the relevant cross sections and the upper limit for the kinetic mixing factor $\varepsilon$ depending on the mass $m_{\Ap}$.
Our calculations of the hidden photon production cross section from electron scattering experiments are compared with the approximate formulae of Ref.~\cite{Bjorken:2009mm} in Sec.~\ref{sec:WW_analysis}.
In Sec.~\ref{sec:future} we present our results for the study of the setups of the DarkLight and HPS-type experiments.
Section~\ref{sec:outlook} presents our conclusions and outlook.
\section{Calculational Method\label{sec:method}}
In this work we denote the four-momenta of the initial and final beam electrons by $k=(E_0,\,\vec{k})$ and $k^\prime=(E_e^\prime,\,\vecp{k})$; the four-momenta of the initial and final target state by $p=(E_p,\,\vec{p})$ and $p^\prime=(E_p^\prime,\,\vecp{p})$ and the lepton pair four-momenta by $l_-=(E_-,\,\vec{l}_-)$ and $l_+=(E_+,\,\vec{l}_+)$, for the lepton and anti-lepton, respectively.
The initial and final electron spins are denoted by $s_k$ and $s_k^\prime$; the spins of the initial and final proton by $s_p$ and $s_p^\prime$; and the spins of the created lepton and anti-lepton by $s_-$ and $s_+$.
Furthermore we follow the conventions of Bjorken and Drell \cite{Bjorken:1964}.
\begin{figure}
\begin{tabular}{cc}
\includegraphics[width=.48\linewidth]{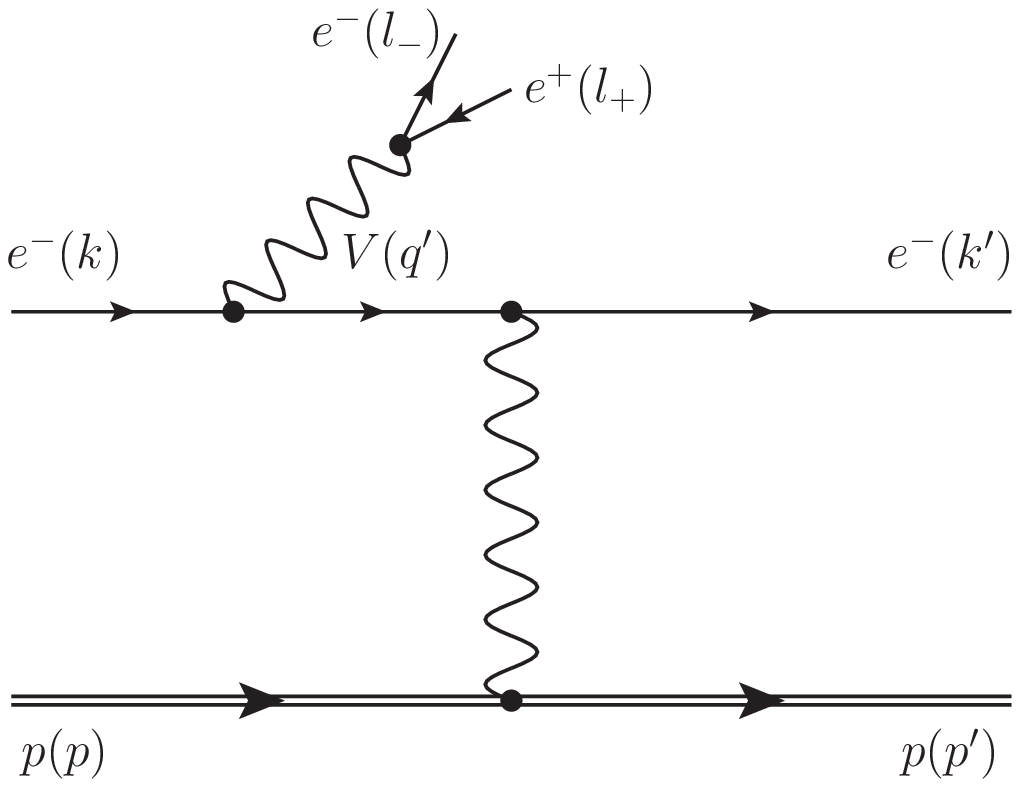}& \includegraphics[width=.48\linewidth]{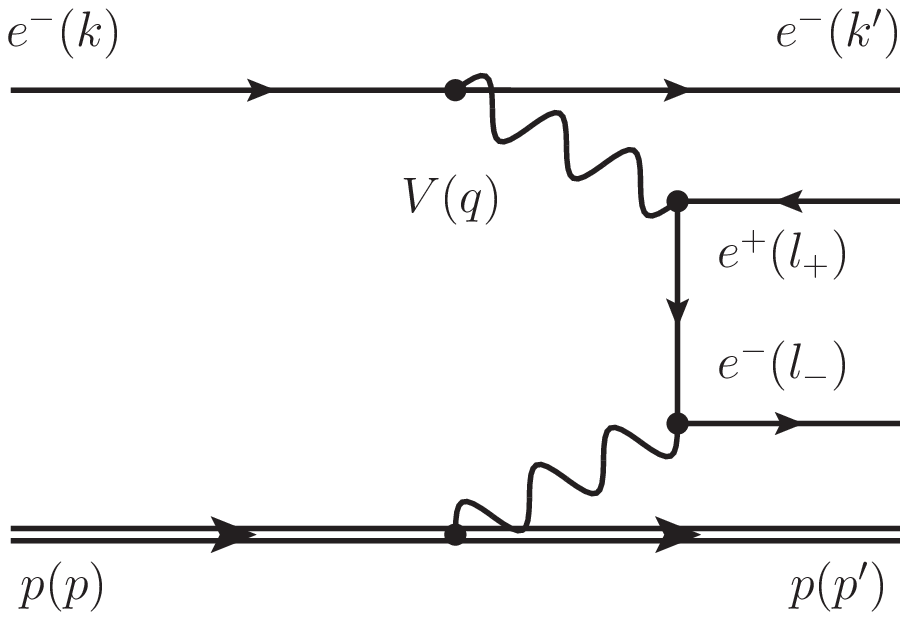}\\[-.3cm]
(TL) & (SL)\\[.3cm]
\multicolumn{2}{c}{\includegraphics[width=.48\linewidth]{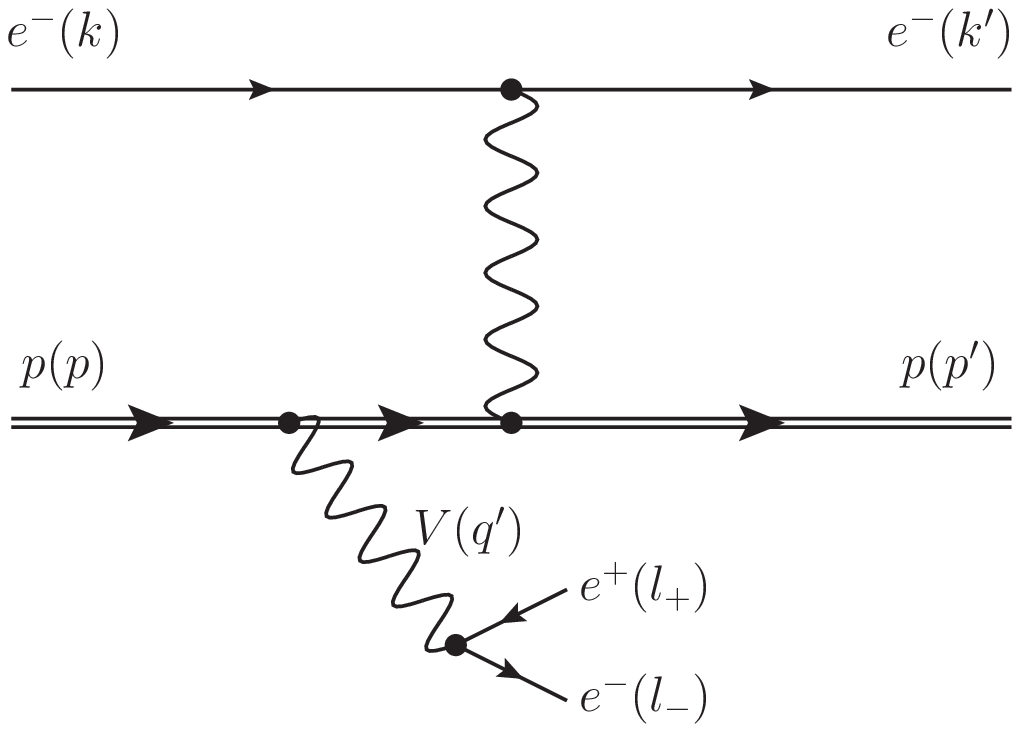}}\\[-.3cm]
\multicolumn{2}{c}{(VVCS)}
\end{tabular}
\caption{Tree level Feynman diagrams contributing to the $e p \rightarrow e p l^+l^-$ amplitude.
The (TL) and (VVCS) diagrams with $e^+e^-$ emission from final lines as well as the (SL) diagrams with $e^+e^-$ lines crossed are also understood.
In addition to these direct (D) diagrams the exchange terms (X), which consist of the same set of diagrams with scattered electron and electron of the $e^+e^-$ pair exchanged, also contribute.
\label{fig:feyn_ep-epll}}
\end{figure}

The signal cross section is described by the Feynman diagrams of Fig.~\ref{fig:feyn_ep-epll} in which the boson $V$ is timelike.
For the production from a heavy nucleus, the radiation off the target is suppressed by its large mass and therefore only the timelike (TL) diagrams contribute.
The emission of a spacelike (SL) hidden photon is strongly suppressed and can be safely neglected.
If a proton is used as target, the doubly virtual Compton scattering (VVCS) off the hadronic state is not suppressed that strongly and therefore we include also the VVCS diagrams into the amplitude of the signal cross section.

For the TL and VVCS diagrams one has for the isolated $\Ap$ production process 
\begin{align}
 \mathcal{M}_{\Ap}^\text{TL} &\quad=\frac{i \,e^4\,\varepsilon^2}{\left(p^\prime-p\right)^2}\,\frac{-g^{\alpha \beta} + {q^{\prime \alpha}q^{\prime \beta} }/{m_{\Ap}^2}}{q^{\prime 2}-m_{\Ap}^2+i\,m_{\Ap}\,\Gamma_{\Ap}}\,J_N^\mu\,\mathcal{I}_{\mu\alpha}\,j^\text{pair}_\beta,\label{epepll:eq_M_ab_Ap-1}
\end{align}
and
\begin{align}
 \mathcal{M}_{\Ap}^\text{VVCS} &\quad=\frac{-i \,e^4\,\varepsilon^2}{q^2}\,\frac{-g^{\alpha \beta} + {q^{\prime \alpha}q^{\prime \beta} }/{m_{\Ap}^2}}{q^{\prime 2}-m_{\Ap}^2+i\,m_{\Ap}\,\Gamma_{\Ap}}\,j_\text{beam}^\mu\,\mathcal{H}_{\mu\alpha}\,j^\text{pair}_\beta \label{epepll:eq_M_vcs_Ap-1},
\end{align}
where $\Gamma_{\Ap}$ denotes the total $\Ap$ decay width.
The total $\Ap$ decay width can be estimated as function of the partial decay width to lepton pairs $\Gamma_{\Ap\rightarrow l^+l^-}$
\[
 \Gamma_\Ap = \Gamma_{\Ap\rightarrow l^+l^-} N_\mathrm{eff},
\]
where $N_\mathrm{eff}$  is a weight to account for other degrees of freedom to SM decays, and the partial decay width to a SM lepton pair $l^+l^-$ is
\[
  \Gamma_{\Ap\rightarrow l^+l^-} = \frac{ \alpha \varepsilon^2}{3 m_{\Ap}^2}\sqrt{m_\Ap^2-4m_l^2}\,(m_\Ap^2+2m_l^2),
\]
with $m_l$ the mass of the decay leptons.
The leptonic and hadronic tensors are given by
\begin{align}
\begin{split}
\mathcal{I}_{\mu\alpha} &= \ubar_e(k^\prime,s_k^\prime) \left( \gamma_\mu \frac{ \fslash{(k-q^\prime)} +m}{\left(k-q^\prime \right)^2-m^2} \gamma_\alpha
+\gamma_\alpha \frac{ \fslash{(k^\prime+ q^\prime)} +m}{\left(k^\prime+q^\prime \right)^2-m^2} \gamma_\mu \right) u_e(k,s_k),\\
\mathcal{H}_{\mu\alpha} &= \ubar_p(p^\prime,s_p^\prime) \left( \Gamma_\mu(q_t + q^\prime) \frac{ \fslash{(p-q^\prime)} +M_N}{\left(p-q^\prime \right)^2-M_N^2} \Gamma_\alpha(-q^\prime) +\Gamma_\alpha(-q^\prime) \frac{ \fslash{(p^\prime+ q^\prime)} +M_N}{\left(p^\prime+q^\prime \right)^2-M_N^2} \Gamma_\mu(q_t + q^\prime) \right) u_p(p,s_p),\label{eq:leptonic_TL_hadronic_VVCS_tensors}
\end{split}
\end{align}
with $m$ ($M_N$) denoting the mass of the electron (hadron).
The leptonic currents read as
\begin{align*}
 j^\text{pair}_\beta &= \ubar_l(l_-,s_-)\, \gamma_\beta\, v_l(l_+,s_+),\\
 j^\text{beam}_\beta &= \ubar_e(k^\prime,s_k^\prime)\, \gamma_\beta\, u_e(k,s_k).
\end{align*}

In the case of a proton target the hadronic current $  J_N^\mu$ is parametrized as
\[
   J_N^\mu =\ubar_N(p^\prime, s_p^\prime)\, \Gamma^\mu\, u_N(p,s_p),
\]
with $\Gamma_\mu (Q_t^2) \equiv F_1(Q_t^2)\, \gamma_\mu + F_2(Q_t^2)\,i\, \sigma_{\mu \nu} {q_t^\nu}/{2M}$ using the Dirac ($F_1$) and Pauli ($F_2$) form factors and $Q_t={-(p-p^\prime)^2}>0$.
Furthermore, we parametrize the form factors $F_1$ and $F_2$ by a linear combination of the electric and magnetic Sachs form factors, for which we use a standard dipole fit in the spacelike as well as timelike regions for low momentum transfer $\left|Q_t^2 \right|\lesssim 1\,\GeV^2$, which is the region of interest in this work.
For a heavy nucleus the hadronic current can be written to good approximation as
\[
 J_N^\mu = Z \cdot F_\mathrm{el}(Q_t)\cdot (p+p^\prime)^\mu,
\]
where 
\[
F_\mathrm{el}(Q_t)={3}/{\left(Q_t\,R\right)^3}\cdot \left({\sin{(Q_t\,R)}} - Q_t R\, \cos{(Q_t\,R)}\right),
\]
is the nuclear charge form factor with $R=1.21\,\mathrm{fm}\cdot A^{\frac{1}{3}}$ \cite{Beranek:2013yqa,Friedrich:1982}.

The amplitude of the QED background is given by the coherent sum over all diagrams shown in Fig.~\ref{fig:feyn_ep-epll}, where each amplitude reads
\begin{align}
  \mathcal{M}_{\gamma^\ast}^\text{TL} &\quad=\frac{i \,e^4}{\left(p^\prime-p\right)^2}\,\frac{-g^{\alpha \beta}}{q^{\prime 2}}\,J_N^\mu\,\mathcal{I}_{\mu\alpha}\,j^\text{pair}_\beta, \label{epepll:eq_M_ab_gamma-1}\\
  \mathcal{M}_{\gamma^\ast}^{SL} &\quad=\frac{i \,e^4}{\left(p^\prime-p\right)^2}\,\frac{-g^{\alpha \beta}}{q^2}\,J_N^\mu\,\tilde{\mathcal{I}}_{\mu\alpha}\,j^\text{beam}_\beta \label{epepll:eq_M_cd_gamma-1},\\
  \mathcal{M}_{\gamma^\ast}^\text{VCS} &\quad=\frac{-i \,e^4}{q^2}\,\frac{-g^{\alpha \beta}}{q^{\prime 2}}\,j_\text{beam}^\mu\,\mathcal{H}_{\mu\alpha}\,j^\text{pair}_\beta, \label{epepll:eq_M_vcs_gamma-1}
\end{align}
with
\begin{align*}
 \tilde{\mathcal{I}}_{\mu\alpha} &=  \ubar_l(l_-,s_-) \left( \gamma_\mu \frac{ \fslash{(q-l_+)} +m_l}{\left(q-l_+ \right)^2-m_l^2 } \gamma_\alpha + \gamma_\alpha \frac{ \fslash{(l_--q)} +m_l}{\left(l_- - q \right)^2-m_l^2} \gamma_\mu \right) v_l(l_+,s_+).
 \end{align*}
In the case, that the species of the beam lepton and of the lepton pair are the same, besides the direct contribution denoted by $D$ given by the amplitudes above, the exchange term ($X$) has to be taken into account, where the two negatively charged leptons are interchanged.
Therefore the full amplitude of the process reads
\begin{align}
 \mathcal{M}_{\Ap+\gamma^\ast} &= \left(\mathcal{M}_{\D,\,\Ap}^\text{TL} + \mathcal{M}_{\D,\,\gamma^\ast}^\text{TL} +\mathcal{M}_{\D,\,\gamma^\ast}^\text{SL} \right) - \left(\mathcal{M}_{\X,\,\Ap}^\text{TL} + \mathcal{M}_{\X,\,\gamma^\ast}^\text{TL} +\mathcal{M}_{\X,\,\gamma^\ast}^\text{SL} \right)
 \label{eq_M_A'+g_full},
\end{align}
and
\begin{align*}
 \mathcal{M}_{\Ap+\gamma^\ast} &= \left(\mathcal{M}_{\D,\,\Ap}^\text{TL} +\mathcal{M}_{\D,\,\Ap}^\text{VVCS} + \mathcal{M}_{\D,\,\gamma^\ast}^\text{TL} +\mathcal{M}_{\D,\,\gamma^\ast}^\text{SL}+\mathcal{M}_{\D,\,\gamma^\ast}^\text{VVCS} \right)- \left(\mathcal{M}_{\X,\,\Ap}^\text{TL} +\mathcal{M}_{\X,\,\Ap}^\text{VVCS} + \mathcal{M}_{\X,\,\gamma^\ast}^\text{TL} +\mathcal{M}_{\X,\,\gamma^\ast}^\text{SL} +\mathcal{M}_{\X,\,\gamma^\ast}^\text{VVCS} \right),
\end{align*}
for a heavy nucleus target and proton target, respectively.

The cross section of the $e (A,\,Z) \rightarrow e (A,\,Z) e^+ e^-$ process is computed from the general expression for $2\rightarrow 4$ particle processes
\begin{align}
\begin{split}
 d \sigma&=\frac{1}{4\sqrt{(k \cdot p)^2-m^2M^2}}\,(2\pi)^4 
\delta^{(4)}\left(k+p-k^\prime-p^\prime-l_--l_+\right)\\
 &\quad\times \frac{d^3\vecp{k}}{(2\pi)^3\,2\, E_e^\prime}\, \frac{d^3\vecp{p}}{(2\pi)^3\,2\,E_p^\prime}\, \frac{d^3\vec{l_-}}{(2\pi)^3\,2\,E_-}\, \frac{d^3\vec{l_+}}{(2\pi)^3\,2\,E_+}\,\overline{\left|\mathcal{M}\right|^2}\label{eq_dcs-gen},
\end{split}
\end{align}
where the explicit expressions of the used cross section formulae are given in Ref.~\cite{Beranek:2013yqa}.

In order to provide realistic predictions for the studied experimental setups we calculate the acceptance integrated cross sections.
To obtain the acceptance integrated cross section $\Delta\sigma$, which can be related to experimental count rates by multiplication with the luminosity, we perform an 8-fold numerical integration of the differential cross section within the corresponding limits of each experiment.
For that purpose we use graphics processing units and the implementation of the VEGAS integration algorithm, as described in \cite{Beranek:2013yqa,Lepage:1977sw,nvidia,Kanzaki:2010ym}.

The pure signal cross section is denoted by
\begin{equation}
 \Delta\sigma_\Ap \propto \left|\mathcal{M}^\text{TL}_{\D,\Ap} \right|^2, \label{eq:Delta_sigma_gammaprime_nucleus}
\end{equation}
and
\begin{equation}
 \Delta\sigma_\Ap \propto \left|\mathcal{M}^\text{TL}_{\D,\Ap} + \mathcal{M}^\text{VVCS}_\D \right|^2, \label{eq:Delta_sigma_gammaprime_proton}
\end{equation}
for the production off a heavy nucleus and a proton, respectively.
Note, that due to the very narrow width entering the $\Ap$ propagator in Eqs.(\ref{epepll:eq_M_ab_Ap-1}) and (\ref{epepll:eq_M_vcs_Ap-1}) the isolated signal cross section does not need to be anti-symmetrized since this contribution is vanishingly small.

The radiative background is described by the acceptance integrated cross sections for a nucleus and proton target
\begin{equation}
 \Delta\sigma_\gamma \propto \left| \mathcal{M}_{\D,\gamma^\ast}^\text{TL} + \mathcal{M}_{\D,\gamma^\ast}^\text{SL} - \mathcal{M}_{\X,\gamma^\ast}^\text{TL} - \mathcal{M}_{\X,\gamma^\ast}^\text{SL} \right|^2, \label{eq:Delta_sigma_gamma_nucleus}
\end{equation}
and
\begin{equation}
 \Delta\sigma_\gamma \propto \left| \mathcal{M}_{\D,\gamma^\ast}^\text{TL} + \mathcal{M}_{\D,\gamma^\ast}^\text{SL} + \mathcal{M}_{\D,\gamma^\ast}^\text{VVCS}- \mathcal{M}_{\X,\gamma^\ast}^\text{TL} - \mathcal{M}_{\X,\gamma^\ast}^\text{SL} - \mathcal{M}_{\X,\gamma^\ast}^\text{VVCS} \right|^2. \label{eq:Delta_sigma_gamma_proton} 
\end{equation}

In our previous work \cite{Beranek:2013yqa} we have used Eq.~(19) of Ref.~\cite{Bjorken:2009mm} to calculate the exclusion limit for $\varepsilon^2$ depending on the hidden photon mass $m_\Ap$.
In this work we calculate the hidden photon cross section directly using the amplitudes given in Eqs.~(\ref{epepll:eq_M_ab_Ap-1}) and (\ref{epepll:eq_M_vcs_Ap-1}).
For that purpose we exploit, that the hidden photon propagator describes a Breit-Wigner distribution.
The integral of the differential cross section over a narrow, but sufficient large invariant mass range of the lepton pair needs to be independent of the kinetic mixing factor $\varepsilon^2$.
Therefore we calculate the signal cross section by setting $\varepsilon=\widetilde{\varepsilon}\equiv10^{-1}$ during the numerical integration, which leads to a reasonable decay width for fixed-target experiments.
Eventually, we divide the signal cross section by $\tilde{\varepsilon}^2$ to normalize it to a cross section of $\varepsilon^2=1$.
We have compared the direct calculation of the hidden photon cross section with the approximation using Eq.~(19) of Ref.~\cite{Bjorken:2009mm} and find, that both methods are in good agreement.

As predicted upper limit for the kinetic mixing factor we find
\begin{align}
 \varepsilon^2 &= \frac{N_\sigma}{\sqrt{\Delta \sigma_\gamma \times L}} \frac{\Delta \sigma_\gamma \times \tilde{\varepsilon}^2}{\Delta \sigma_\Ap (\varepsilon=\tilde{\varepsilon}^2)},\label{eq:exclusionlimit}
\end{align}
where $L$ denotes the integrated luminosity obtained in the particular experiment and $N_\sigma$ accounts for the fact, that the exclusion limit is valid at the $(N_\sigma \times \sigma)$ level.
To compare with other publications, we use $N_\sigma = 2$, if not explicitly mentioned.
Note, that in the calculation of exclusion limits from experimental data the estimate $\sqrt{\Delta \sigma_\gamma \times L}$ needs to be replaced by the actual limit found in the analysis of the peak search.
\section{Applicability of the Weizs\"acker-Williams Approximation\label{sec:WW_analysis}}
In Ref.~\cite{Bjorken:2009mm} the Weizs\"acker-Williams (WW) method is used to estimate the cross section of $\Ap$ bremsstrahlung emission induced from the interaction of an electron beam with an atomic target.
An approximate expression for the cross section of the process $e (A,\,Z) \rightarrow e \Ap X$ is found in the framework of the generalized WW approximation
\begin{align}
 d\sigma (e Z \rightarrow e \Ap X)_\text{WW} \propto d\sigma (e \gamma \rightarrow e \Ap) \times \frac{\alpha}{\pi}\,\chi,
\end{align}
where $d\sigma (e \gamma \rightarrow e \Ap)$ is the cross section of the process $e \gamma \rightarrow e \Ap$ and $({\alpha}/{\pi}\,\chi)$ is the effective pseudo-photon flux accounting for the hadronic interaction.

In addition to Ref.~\cite{Bjorken:2009mm}, the WW approximation of the $\Ap$ production cross section was investigated in Ref.~\cite{Andreas:2012mt} in order to extract limits on the $\Ap$ parameter space from past beam dump experiments.
In these publications the differential cross section for $\Ap$ production is given as
\begin{align}
 \frac{d\sigma_\mathrm{WW}}{dx\,d\cos \theta_\Ap} &= 8\alpha^3 \varepsilon^2 E_0^2 x \chi \sqrt{1-\frac{m_\Ap^2}{E_0^2}} \left(\frac{1-x+{x^2}/{2}}{U^2} + \frac{\left(1-x \right)^2 m_\Ap^2}{U^4} - \frac{\left(1-x\right)x m_\Ap^2}{U^3} \right),\label{eq:epepV_WW_1}
\end{align}
where $x={E_\Ap}/{E_0}$ is the fraction of the energies carried by the $\Ap$ and the incident electron and $\theta_\Ap$ is the polar emission angle of the $\Ap$ in the lab frame with respect to the $z$-axis, which corresponds to the beam axis.
Furthermore,
\[
 \chi = \chi \left(E_0,\,m_\Ap,\,Z,\,A \right) = \int_{t_\text{min}}^{t_\text{max}} dt \frac{t-{t_\text{min}}}{t^2}\,G_2 (t),
\]
is the effective photon flux for an atomic nucleus of atomic number $Z$ and particle number $A$, $t=-\Delta^2=-(p^\prime-p)^2$ is the negative of the squared momentum transfer, $t_\text{min}=({m_\Ap^2}/{(2 E_0)})$, $t_\text{max}=m_\Ap^2$.
The form factor $G_2 (t)$ is given in Ref.~\cite{Bjorken:2009mm}.
The function
\[
 U=U\left(x,\,E_0,\,m_\Ap,\,\theta_\Ap \right) = E_0^2 x \theta_\Ap^2 + m_\Ap^2 \frac{1-x}{x}+m^2x
\]
parametrizes the virtuality of the electron in the intermediate state in the Feynman diagram labeled by (TL) of Fig~\ref{fig:feyn_ep-epll}.
This approximate expression for the cross section is valid for large beam energies compared to the mass of the $\Ap$
\[
 m\ll m_\Ap \ll E_0
\]
and small $\Ap$ emission angles
\[
 x\theta_\Ap^2\ll 1.
\]

We compare the results found in the WW approximation with the cross section calculated in leading order of perturbation theory without neglecting the recoil on the hadronic target.
Therefore we calculate the cross section of the process $e(A,\,Z)\rightarrow e\gamma^\prime (A,\,Z)$ for tantalum as nuclear target, which has been used in the experiments performed so far \cite{Merkel:2011ze,Abrahamyan:2011gv}, i.e. $Z=73$, $A=181$, $M\simeq 168\,\GeV$.
Since the inelastic contribution to the cross section is suppressed by $Z$ compared to the elastic contribution \cite{Bjorken:2009mm}, the cross section is underestimated only by less than $3\,\%$ for the considered nuclear target and beam energies, which we have proven numerically.
Hence we investigate the coherent scattering off a nucleus $e(A,\,Z)\rightarrow e\gamma^\prime (A,\,Z)$ instead of the inelastic cross section $e(A,\,Z)\rightarrow e\gamma^\prime X$.
Note, that due to the choice of form factors in Ref.~\cite{Bjorken:2009mm}, the inelastic cross section and the scattering off a spinless hadronic charge distribution as considered in this work lead to the same cross section formulae when restricted to the kinematics in which the break-up of the target is neglected.
The invariant amplitude of the process $eZ\rightarrow e\gamma^\prime Z$ is given by the expression
\[
 \mathcal{M}_{\Ap} =\frac{-i \,e^3\,\varepsilon}{\left(p^\prime-p\right)^2}\,\frac{-g^{\alpha \beta} + {q^{\prime \alpha}q^{\prime \beta} }/{m_{\Ap}^2}}{q^{\prime 2}-m_{\Ap}^2+i\,m_{\Ap}\,\Gamma_{\Ap}}\,J_N^\mu\,\mathcal{I}_{\mu\alpha}\,\varepsilon^\ast_\beta,\label{eq:M_Ap_2-3}
\]
where $\mathcal{I}_{\mu\alpha}$ is given in Eq.~(\ref{eq:leptonic_TL_hadronic_VVCS_tensors}), $J_N^\mu$ denotes the nuclear current and $\varepsilon^\ast_\beta$ is the polarization vector of the $\Ap.$
Due to gauge invariance one has $\mathcal{I}_{\mu\alpha}q^{\prime \alpha}=0$.

The differential cross section in the lab frame reads
\begin{align}
 \frac{d\sigma}{dE_{q^\prime}\,d\Omega_{q^\prime}\,d\Omega_{k^\prime}} &= \frac{1}{32 \Abs{k} M_N} \frac{1}{(2\pi)^5} \frac{\Absp{k}^2\,\Absp{q}}{E_{k^\prime}\,E_{p^\prime}} \,\left| \frac{\Absp{k}}{E_{k^\prime}} - \frac{\Absp{k}-\left(\vec{k}-\vecp{q} \right)\hat{k}^\prime}{E_{p^\prime}} \right|^{-1} \,\overline{\left|\mathcal{M}\right|^2},\label{eq:epepV_dxs_lab-final}
\end{align}
where $\hat{k}^\prime = \vecp{k}/\Absp{k}$ and $\Absp{k}$ is fixed by the energy conserving $\delta$-function as given in Eq.~(A6) of Ref.~\cite{Beranek:2013yqa}.
To compare with the formulae in the WW approximation we integrate Eq.~(\ref{eq:epepV_dxs_lab-final}) over the full solid angle of the scattered electron and the azimuthal angle of the hidden photon.

\begin{figure}[tbp]
\begin{tabular}{cc}
 \includegraphics[angle=-90,width=.48\linewidth]{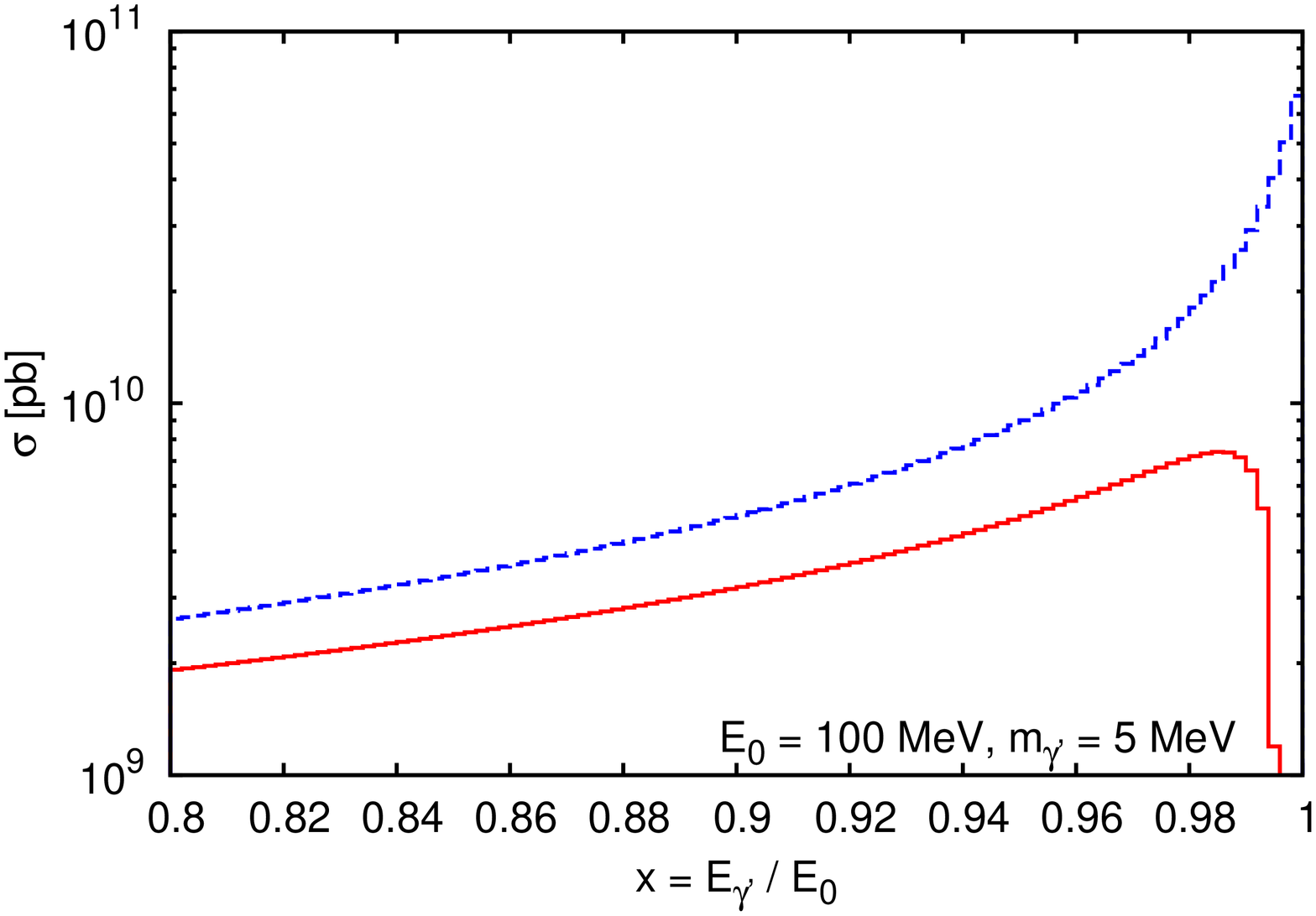}&\includegraphics[angle=-90,width=.48\linewidth]{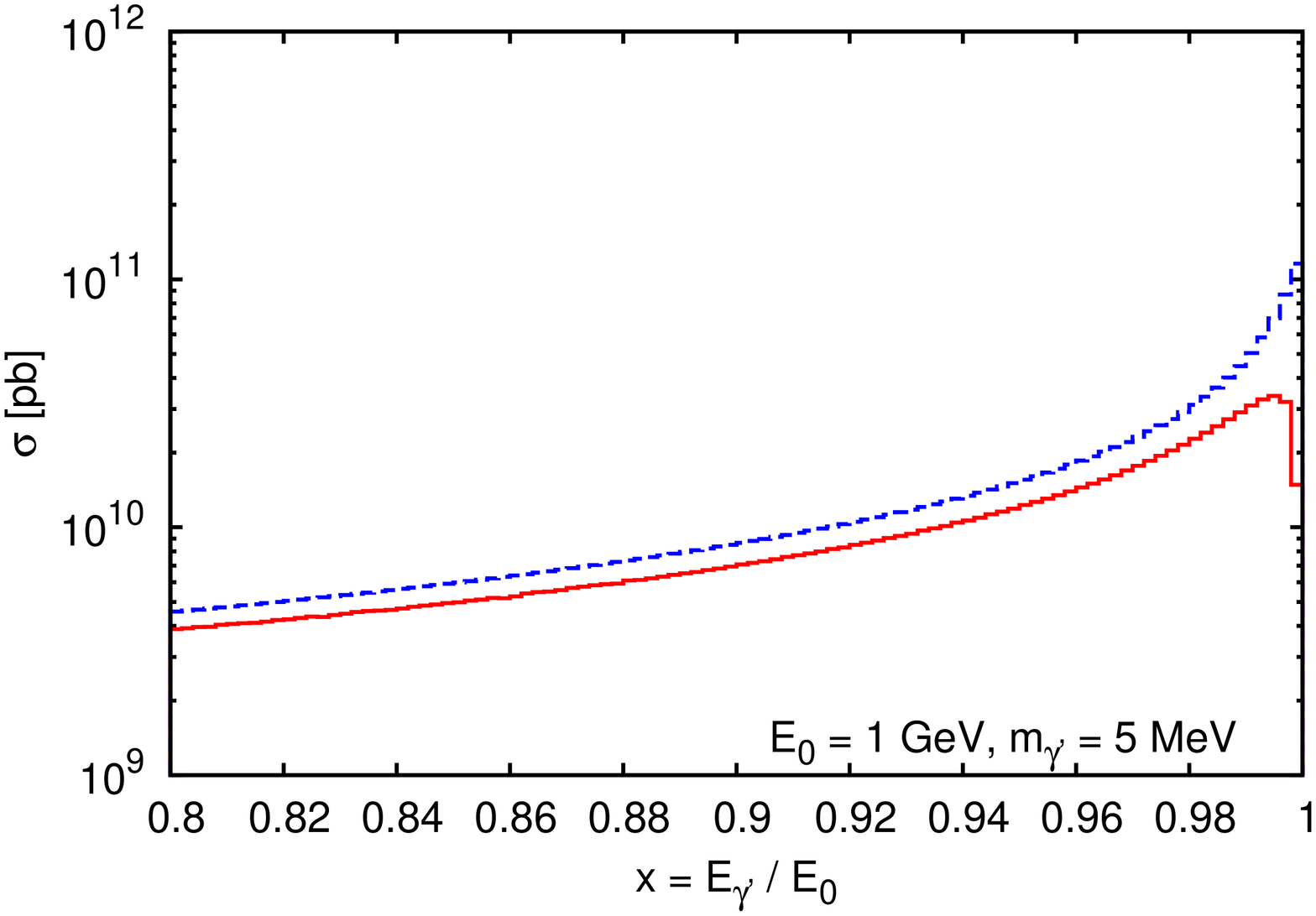}
\end{tabular}
 \caption{Comparison of exact calculation (solid curve) and WW approximation (dashed) for $m_\Ap=5\,\MeV$. The left (right) panel shows the calculation for a beam energy of $E_0=100\,\MeV$ ($1\,\GeV$). For simplicity $\varepsilon^2=1$ is used.\label{fig:epepV_WW_results_5MeV}}
 \end{figure}
 \begin{figure}[htbp]
\begin{tabular}{cc}
  \includegraphics[angle=-90,width=.48\linewidth]{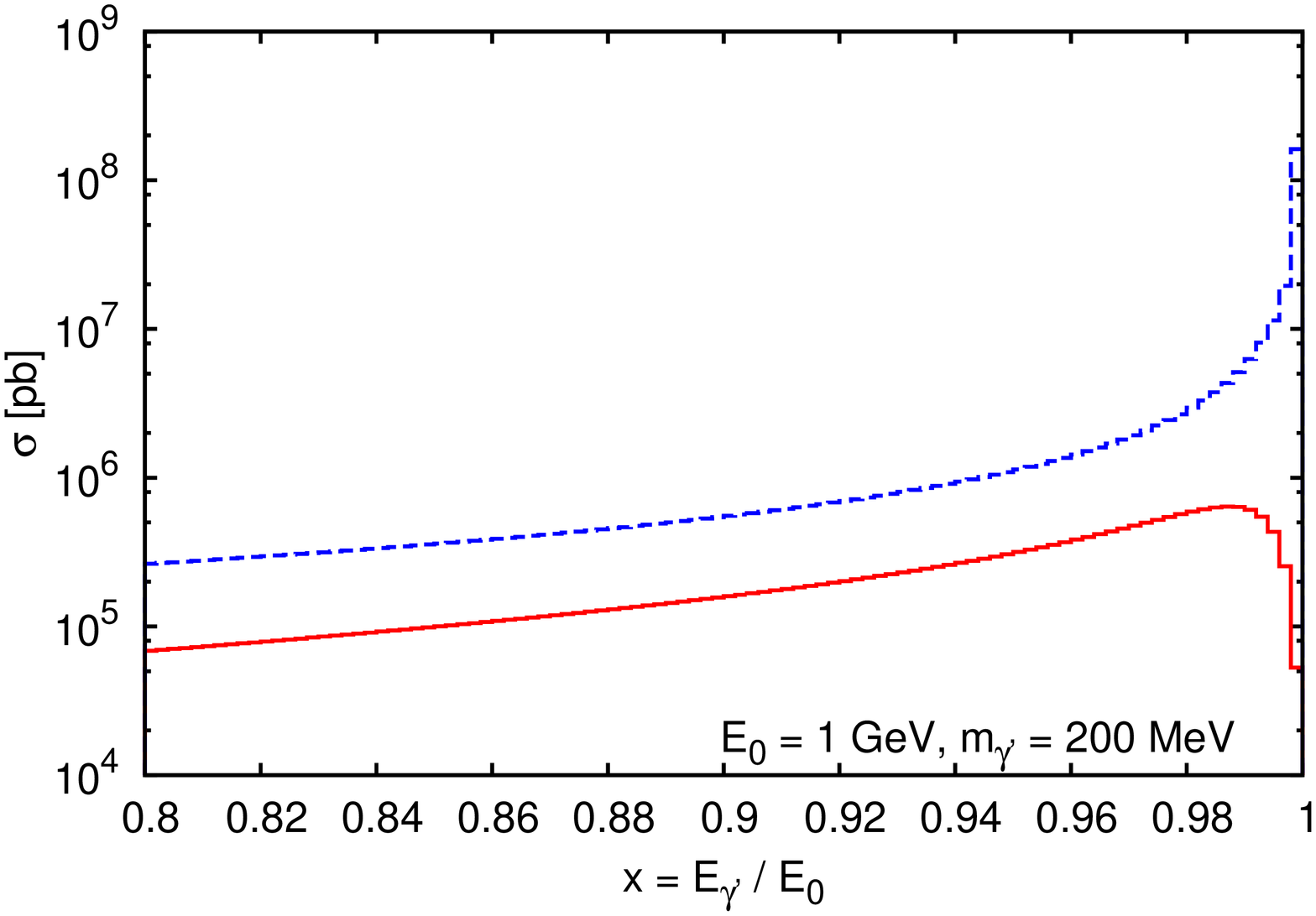}&\includegraphics[angle=-90,width=.48\linewidth]{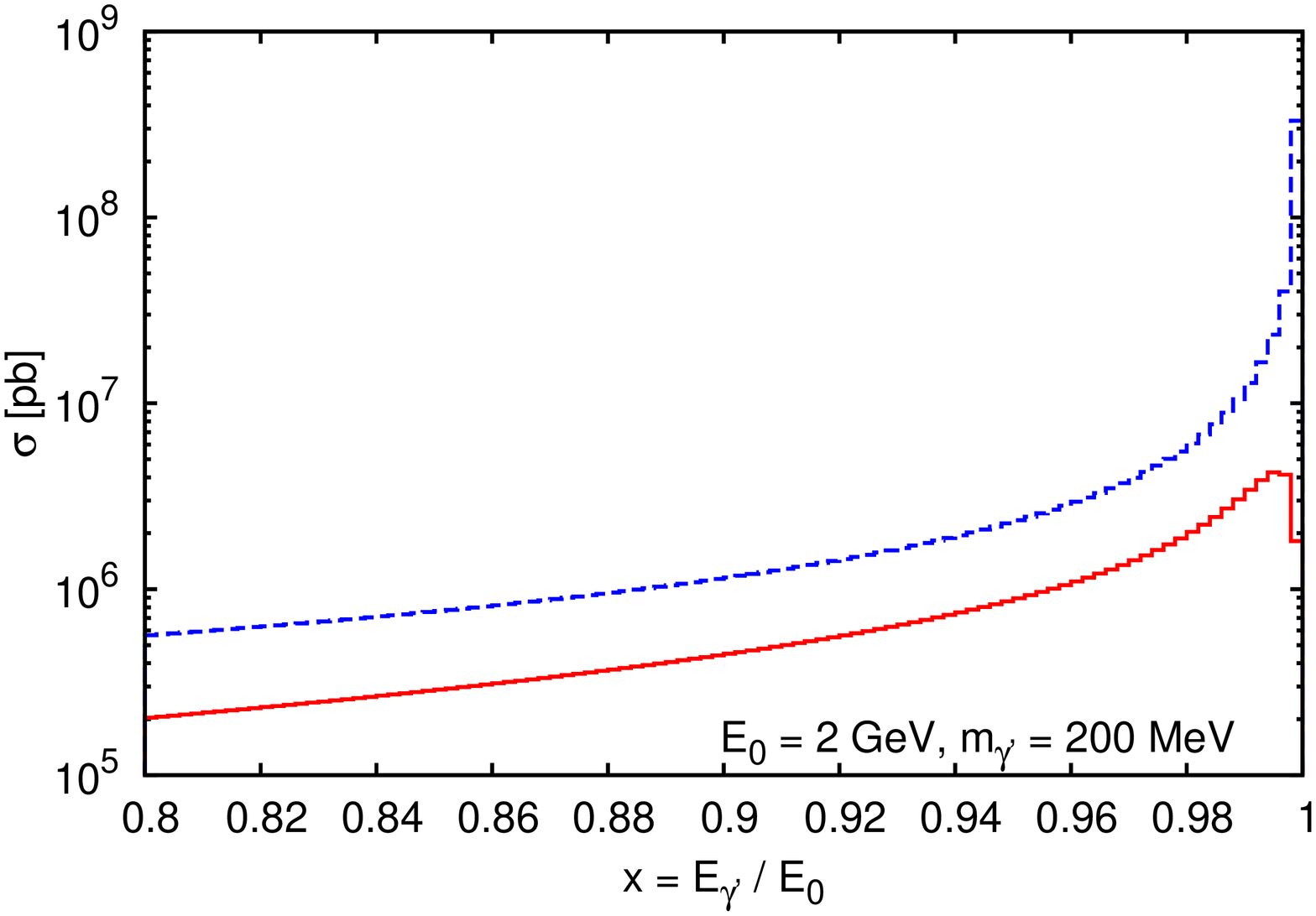}\\
 \includegraphics[angle=-90,width=.48\linewidth]{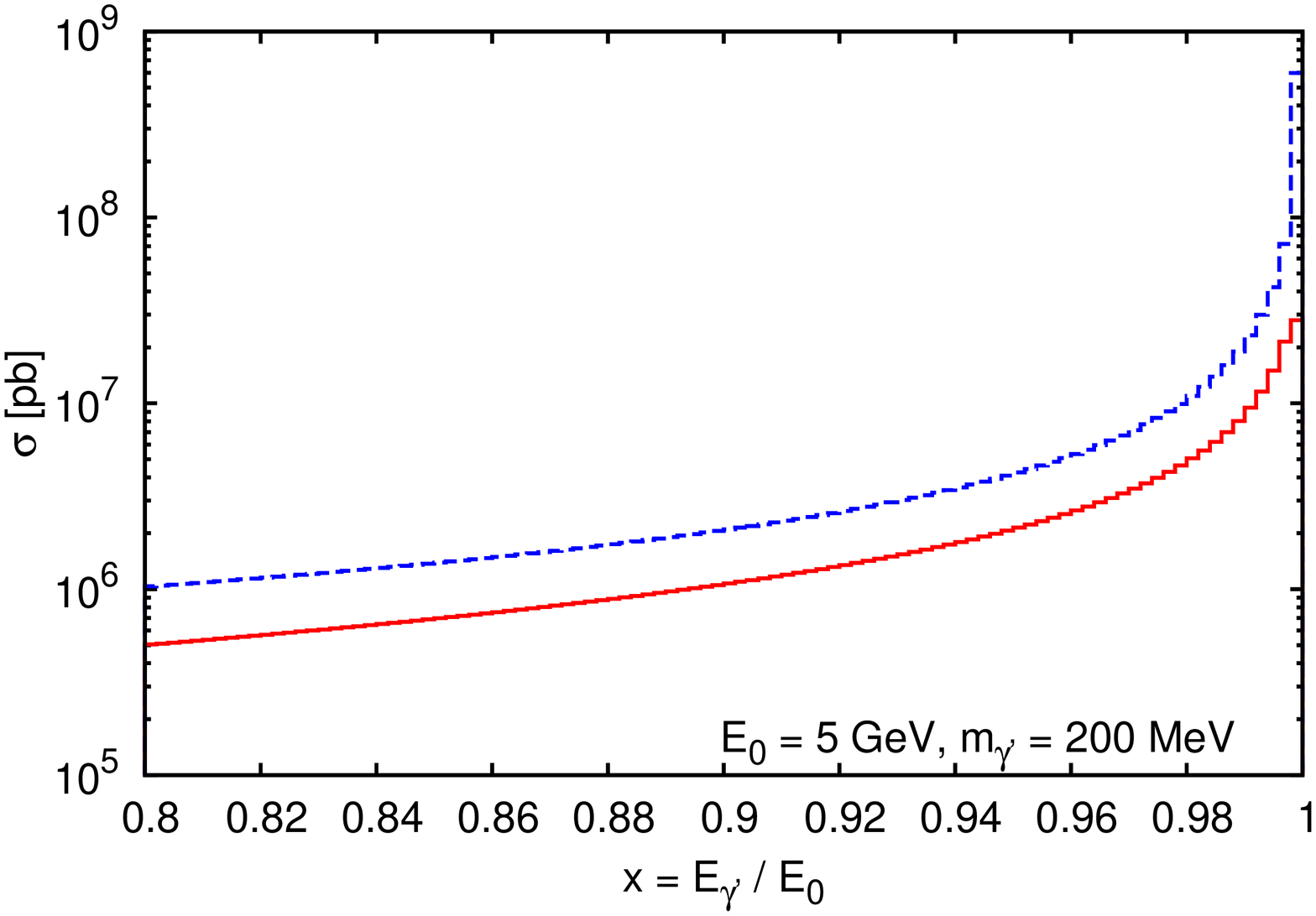}&\includegraphics[angle=-90,width=.48\linewidth]{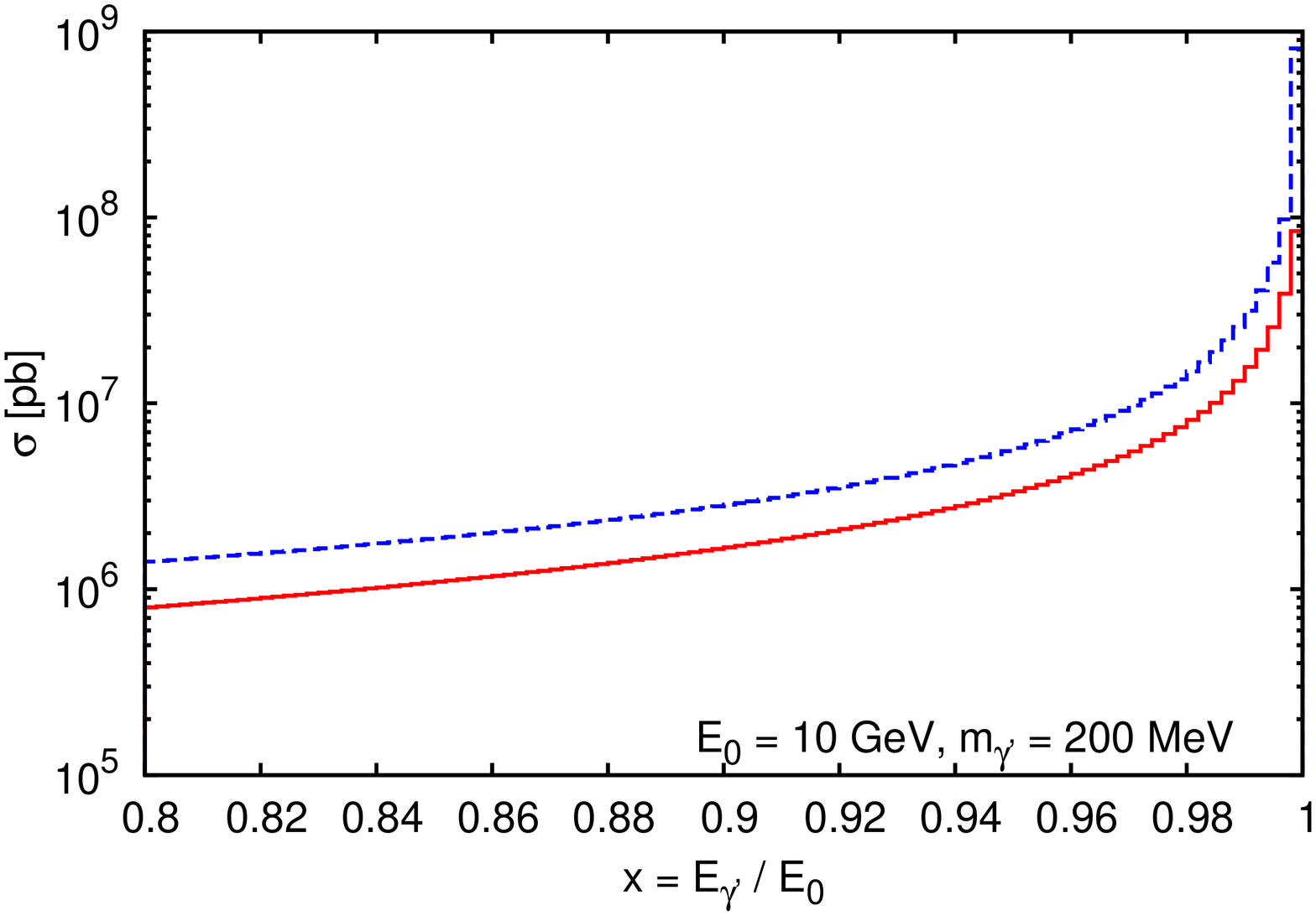}
\end{tabular}
 \caption{Comparison of exact calculation (solid curve) and WW approximation (dashed) for $m_\Ap=200\,\MeV$. The calculation was performed for beam energies from $E_0=1\,\GeV$ (upper left panel) to $E_0=10\,\GeV$ (lower right panel). As before, for simplicity $\varepsilon^2=1$ is used.\label{fig:epepV_WW_results_200MeV}}
 \end{figure}
Although a comparison on the level of the differential cross sections is possible, the acceptance integrated cross sections are compared, since these are the quantities which will be observed in experiments.
This allows for taking all of the possible phase space into account and thus one avoids influence from kinematical cuts on the result.
Furthermore, the integrated cross section can be compared directly to possible experimental data.
For this purpose, the differential cross section in the WW approximation is integrated according to
\begin{align}
  \Delta \sigma &= \int_{x_0}^{x_0+\delta x} dx  \int_{0\,\text{rad}}^{0.5\,\text{rad}} d\theta_\Ap \sin \theta_\Ap \frac{d\sigma}{dx\,\theta_\Ap}, \label{eq:epepV_WW_int1 2}
\end{align}
where the integration over $x$ is performed over an arbitrary small bin with width $\delta x$, where we have chosen $\delta x = 0.002$.
Furthermore, $\theta_\Ap \leq 0.5\,\mathrm{rad}$ was chosen in agreement with Refs.~\cite{Bjorken:2009mm,Andreas:2012mt}, which corresponds to the $\Ap$ forward emission kinematics, in which the WW approximation is expected to be valid.

For simplicity, the atomic form factors of Ref.~\cite{Bjorken:2009mm} were neglected.
Instead, we replace the square of the single elastic form factor by a linear combination of the elastic and quasi-elastic form factors given in Ref.~\cite{Bjorken:2009mm}
\[
 Z^2\, F^2(t) \rightarrow \underbrace{\left(\frac{1}{1+\frac{t}{d}}\right)^2 Z^2}_{G_{\mathrm{el}}} + \underbrace{\left(\frac{1+\frac{t}{(4 M_N^2)}(\mu_p^2-1)}{\left(1+\frac{t^2}{\Lambda^2}\right)^4} \right)^2 Z}_{G_{\mathrm{inel}}},
\]
with $d=0.164\,\GeV^2 \,A^{-2/3}$, and thus
\[
 \chi = \int_{t_\text{min}}^{t_\text{max}} \frac{dt}{t^2} \,\left(G_{\mathrm{el}}(t)+G_{\mathrm{inel}}(t)\right).
\]

The comparison of the calculation in leading order of perturbation theory and the result in the WW approximation is done for a light hidden photon with $m_\Ap=5\,\MeV$ and for $m_\Ap=200\,\MeV$, which is in the typical range of probed masses at MAMI and JLAB \cite{Merkel:2011ze,Abrahamyan:2011gv,HPS}.
The calculations for $m_\Ap=5\,\MeV$ were performed for two different beam energies which were chosen as $E_0=100\,\MeV$ and $E_0=1\,\GeV$.
The results of the this calculation are presented in Fig.~\ref{fig:epepV_WW_results_5MeV}.
The solid curve depicts the result of the exact calculation while the result obtained using the WW approximation is given by the dashed curve.
It turns out from the curves in the left panel of Fig.~\ref{fig:epepV_WW_results_5MeV}, that for $m_\Ap=5\,\MeV$ and $E_0=100\,\MeV$ both calculations show strong deviations.
While in the WW approximation the cross section is strongly increasing for $x\simeq 1$, the cross section calculated in leading order of perturbation theory in this work sharply drops.
Furthermore, the normalizations of the cross sections differ by a factor of 2.
However, for $E_0=1\,\GeV$ both calculations agree within a few per cent except for the region around $x\rightarrow 1$, which is shown in the right panel of Fig.~\ref{fig:epepV_WW_results_5MeV}.

Figure~\ref{fig:epepV_WW_results_200MeV} shows this comparison for $m_\Ap=200\,\MeV$.
One notices, that both methods are also in good agreement for a beam energy which is much larger than $m_\Ap$.
For experiments performed at JLAB, beam energies larger than $2\,\GeV$ are possible, whereas at MAMI the beam energy is below $1.6\,\GeV$.
Therefore, the settings with $E_0=1\,\GeV$ and $E_0=2\,\GeV$ correspond to experiments which can be performed at MAMI, whereas $E_0=2\,\GeV$ to $E_0=10\,\GeV$ correspond to possible experiments at JLAB.

For beam energies of $E_0 \geq 5\,\GeV$ or larger the calculation of this work as well as the WW approximation lead to cross sections which are in good agreement for $x$ close to 1.
However, for a lower beam energy, such as $E_0 = 1\,\GeV$ and $E_0 = 2\,\GeV$, which are shown in the upper panel of Fig.~\ref{fig:epepV_WW_results_200MeV}, the shape of the cross section as well as its normalization significantly differ.
Again, while the cross section in the WW approximation is peaking for $x\simeq 1$, the exact calculation sharply falls off.

The overestimate of the cross section in the WW approximation motivated us to investigate the origin of the deviation.
For that purpose we have re-evaluated the formulae given in Ref.~\cite{Bjorken:2009mm}.
We find, that the assumption, that the minimal momentum transfer does not depend on $x$ and the emission angle $\theta_\Ap$ \cite{Bjorken:2009mm,Essig:2010xa,Andreas:2012mt}, leads to the observed overestimate of the cross section of roughly $30\%$ for large beam energies, where one expects both calculations to be in good agreement.
Therefore, we use Eq.~(A6) of Ref.~\cite{Bjorken:2009mm} as lower limit of the $t$-integration, i.e.
\[
 t_\mathrm{min} = t_\mathrm{min}\left(x,\,\cos{\theta_\Ap}\right) = \left(\frac{U}{2E_0 (1-x)}\right)^2.
\]
The dependence on $x$ and $\theta_\Ap$ causes, that the $t$-integral cannot be evaluated independently from the remaining integration over $x$- and $\cos{\theta_\Ap}$.
\begin{figure}[htb]
 \begin{tabular}{cc}
  \includegraphics[angle=-90,width=.48\linewidth]{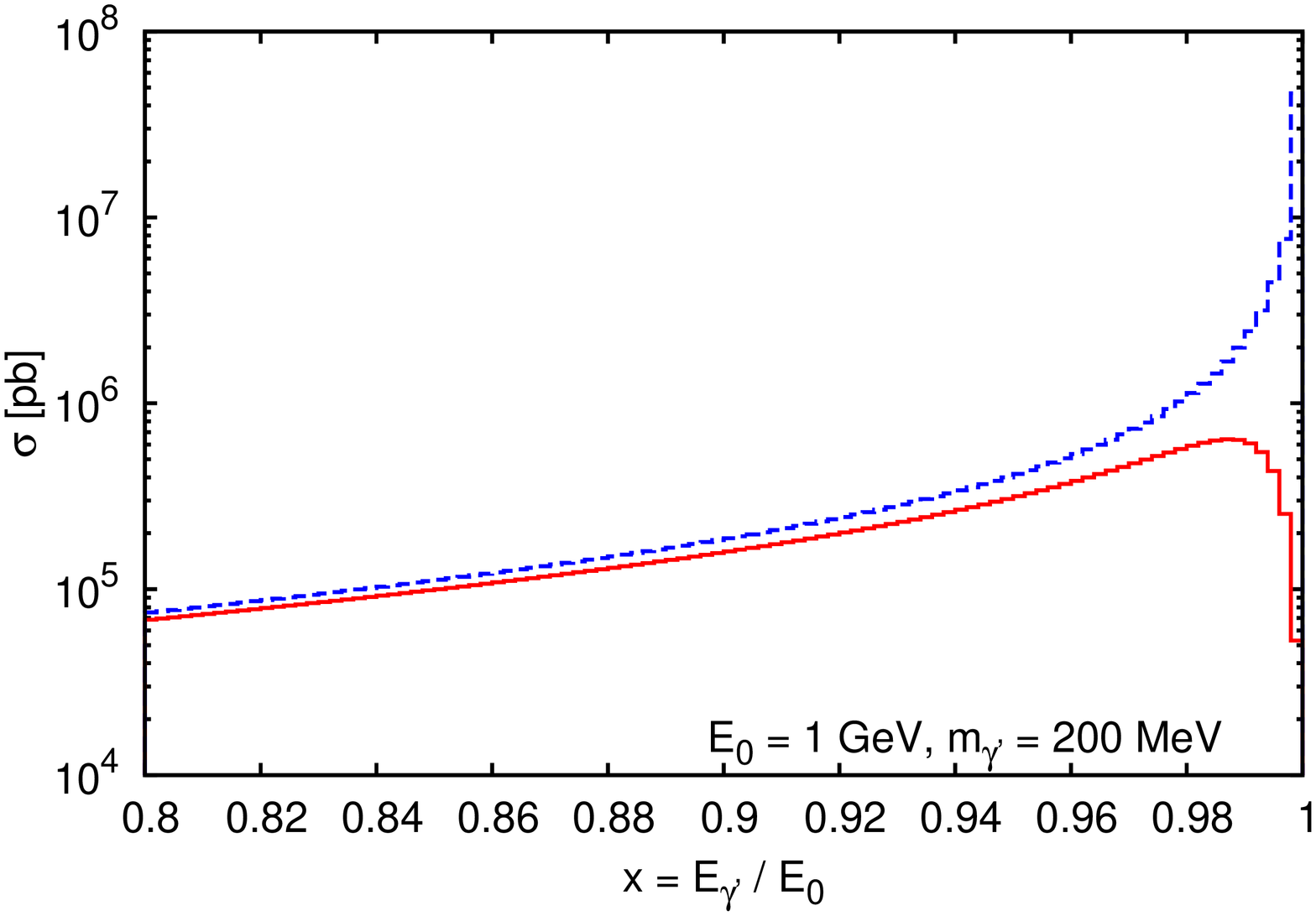}&
  \includegraphics[angle=-90,width=.48\linewidth]{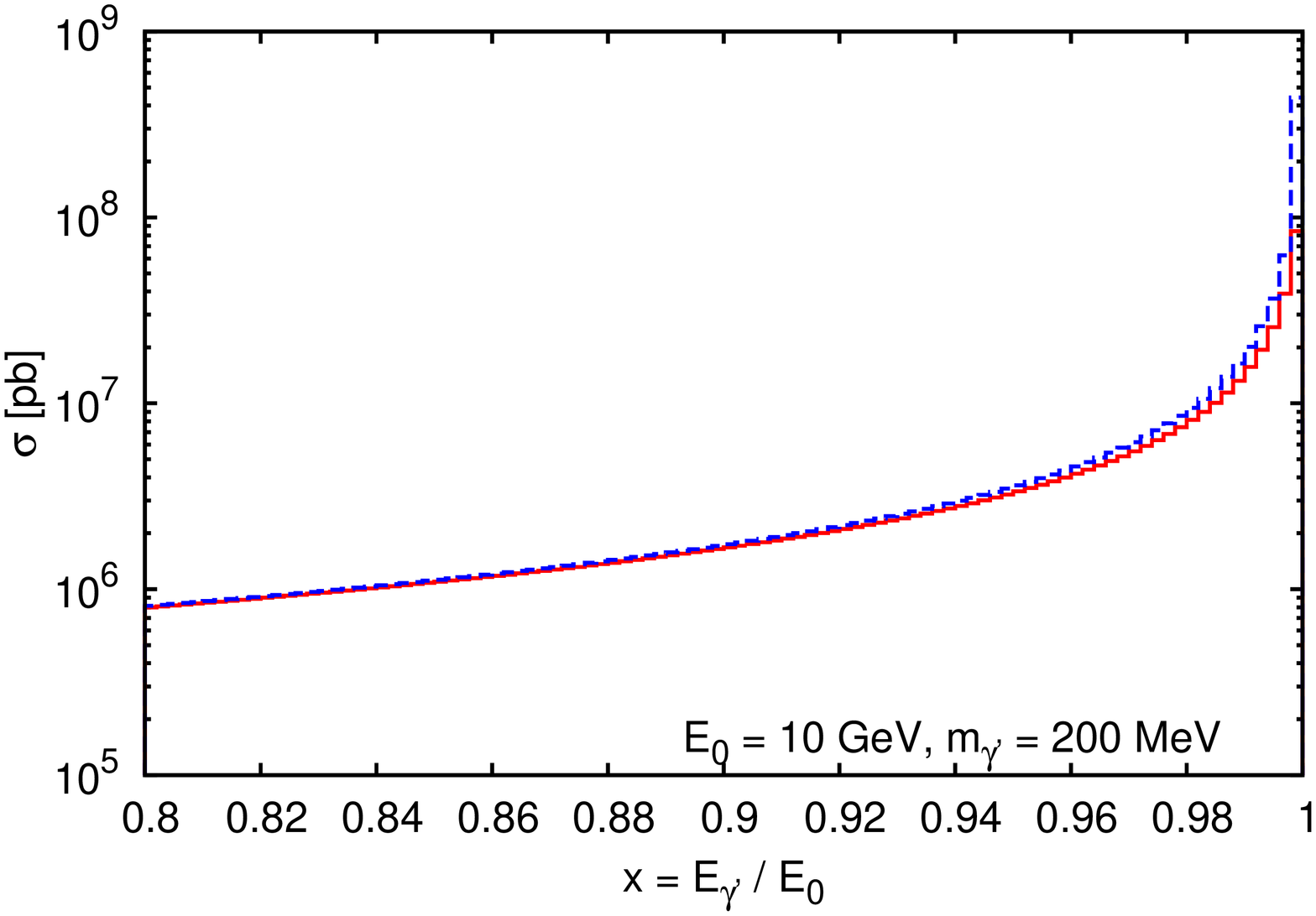}
 \end{tabular}
 \caption{Comparison of the calculation in leading order of perturbation theory (solid curve) and WW approximation, where the dependence of the $t$-integration on $x$- and $\cos{\theta_\Ap}$ is kept (dashed) for $m_\Ap=200\,\MeV$ and $E_0=1\,\GeV$ ($E_0=10\,\GeV$) in the left (right) panel. As before, for simplicity $\varepsilon^2=1$ is used.\label{fig:epepV_WW_results_200MeV-2}}\end{figure}
Performing this more complicated numerical integration, we are able to find an agreement within a few percent between the WW approximation and our calculation, which is presented in Fig.~\ref{fig:epepV_WW_results_200MeV-2}.
Obviously, for $E_0=10\,\GeV$ both calculations are in very good agreement over the full considered $x$-range, which can be seen from the right panel.
However, for $E_0=1\,\GeV$ one still sees a strong deviation for $x\rightarrow 1$.
This effect can be explained by the negligence of the finite momentum transfer to the hadronic state in the WW approximation.

\begin{figure}[htb]
 \begin{center}
  \includegraphics[angle=-90,width=.49\linewidth]{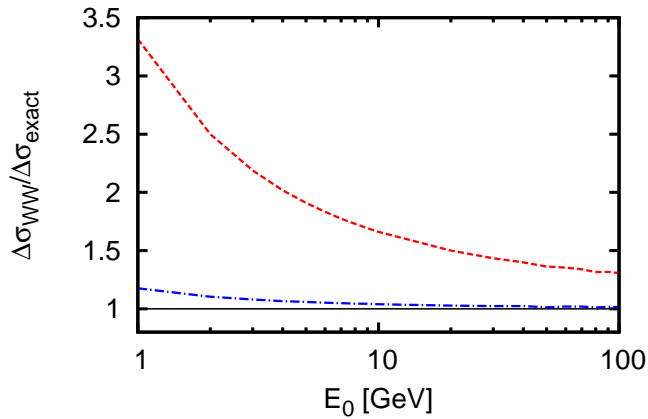}
 \end{center}
\caption{Ratio of the cross section obtained in the WW approximation and in the calculation of this work at $x=E_\Ap/E_0=0.9$. The dashed (dashed-dotted) curve shows the result obtained without (with) $x$- and angular dependence of the minimal momentum transfer $t_\mathrm{min}$.\label{fig:epepV_WW_E0_x0p9}}
\end{figure}
In Fig.~\ref{fig:epepV_WW_E0_x0p9} the ratio of the cross sections calculated with the method of this work and within the WW approximation is plotted as function of the beam energy at fixed $x=E_\Ap/E_0=0.9$. 
Figure~\ref{fig:epepV_WW_E0_x0p9} illustrates, that the deviation between the two calculations which arises from neglecting the $x$- and $\cos{\theta_\Ap}$ dependence of the $t$-integration in case of the WW calculation converges to an overestimate of the cross section by the WW method of around $30\%$.
Taking into account, that the lower limit of the $t$-integral depends on $x$- and ${\theta_\Ap}$, both calculations perfectly agree for large beam energies.

This result is of high importance, since the kind of experiments proposed in Ref.~\cite{Bjorken:2009mm} are planned in such a way, that all of the beam energy is transferred to the $\Ap$.
\section{Calculation of projected exclusion limits for future experiments at JLAB\label{sec:future}}
\subsection{Predictions for a \texorpdfstring{4$\pi$}{4 pi} detector setup (DarkLight)\label{sec:darklight}}
The DarkLight experiment at JLAB plans to investigate the low mass $\Ap$ parameter region.
The experiment is planned at the JLAB Free Electron Laser, which allows for a beam energy of $E_0=100\,\MeV$ with a very high intensity.
As target a hydrogen gas target shall be used.
The goal is to acquire an integrated luminosity of $1\,\mathrm{ab}^{-1}$.

In our previous work we have already studied the feasibility of a possible experiment using two high-resolution spectrometers detecting the lepton pair in forward direction in the context of a feasibility study for the new MESA facility at Mainz.
The DarkLight detector covering nearly the full solid angle is intended to detect all particles involved in the reaction.
Since the beam energy $E_0=100\,\MeV$ is below the pion threshold, the signal and background process are described by Eqs.~(\ref{eq:Delta_sigma_gammaprime_proton}) and (\ref{eq:Delta_sigma_gamma_proton}) without any further assumptions.

The kinematical limits accounting for the detector geometry used are written in Tab.~\ref{tab:DarkLight_kin}.
\begin{table}[ht]
 \begin{tabular}{c||c|c}
 Quantity & Setting I & Setting II\\
 \hline
 \hline
 $\Abs{l} = \Absp{k},\,\Abs{l_-},\,\Abs{l_+}$ & $\Abs{l}\geq5\,\MeV$ & $\Abs{l}\geq 5\,\MeV$ \\[2pt]
 $\theta_{l} = \theta_{k^\prime},\,\theta_{l_-},\,\theta_{l_+}$ & $25^\circ \leq \theta_{l} \leq 155^\circ$ & $20^\circ \leq \theta_{l} \leq 160^\circ$\\[2pt]
 $\phi_{l} = \phi_{k^\prime},\,\phi_{l_-},\,\phi_{l_+}$ & $0^\circ \leq \phi_{l} \leq 360^\circ$ & $0^\circ \leq \phi_{l} \leq 360^\circ$\\[2pt]
 \hline
 $\Absp{p}$ & $\Absp{p}\geq 2\,\MeV$ & $\Absp{p}\geq 2\,\MeV$ \\[2pt]
 $\theta_{p^\prime}$ & $5^\circ \leq \theta_{p^\prime} \leq 175^\circ$ & $4^\circ \leq \theta_{p^\prime} \leq 176^\circ$ \\[2pt]
 $\phi_{p^\prime}$ & $0^\circ \leq \phi_{p^\prime} \leq 360^\circ$ & $0^\circ \leq \phi_{p^\prime} \leq 360^\circ$
\end{tabular}
\caption{Choice of kinematical limits used within the calculations for DarkLight.\label{tab:DarkLight_kin}}
\end{table}
Our simulations were done with two slightly different settings:
Setting I corresponds to the one discussed in the proposal of the DarkLight experiment \cite{DarkLight}.
In Setting II, the range of the polar angles of the final state proton and leptons is slightly extended, which allows for the acceptance of leptons emitted more closely to the beam axis \cite{Corliss:private}.

\begin{figure*}[htbp]
 \begin{tabular}{cc}
  \includegraphics[angle=-90,width=.49 \linewidth]{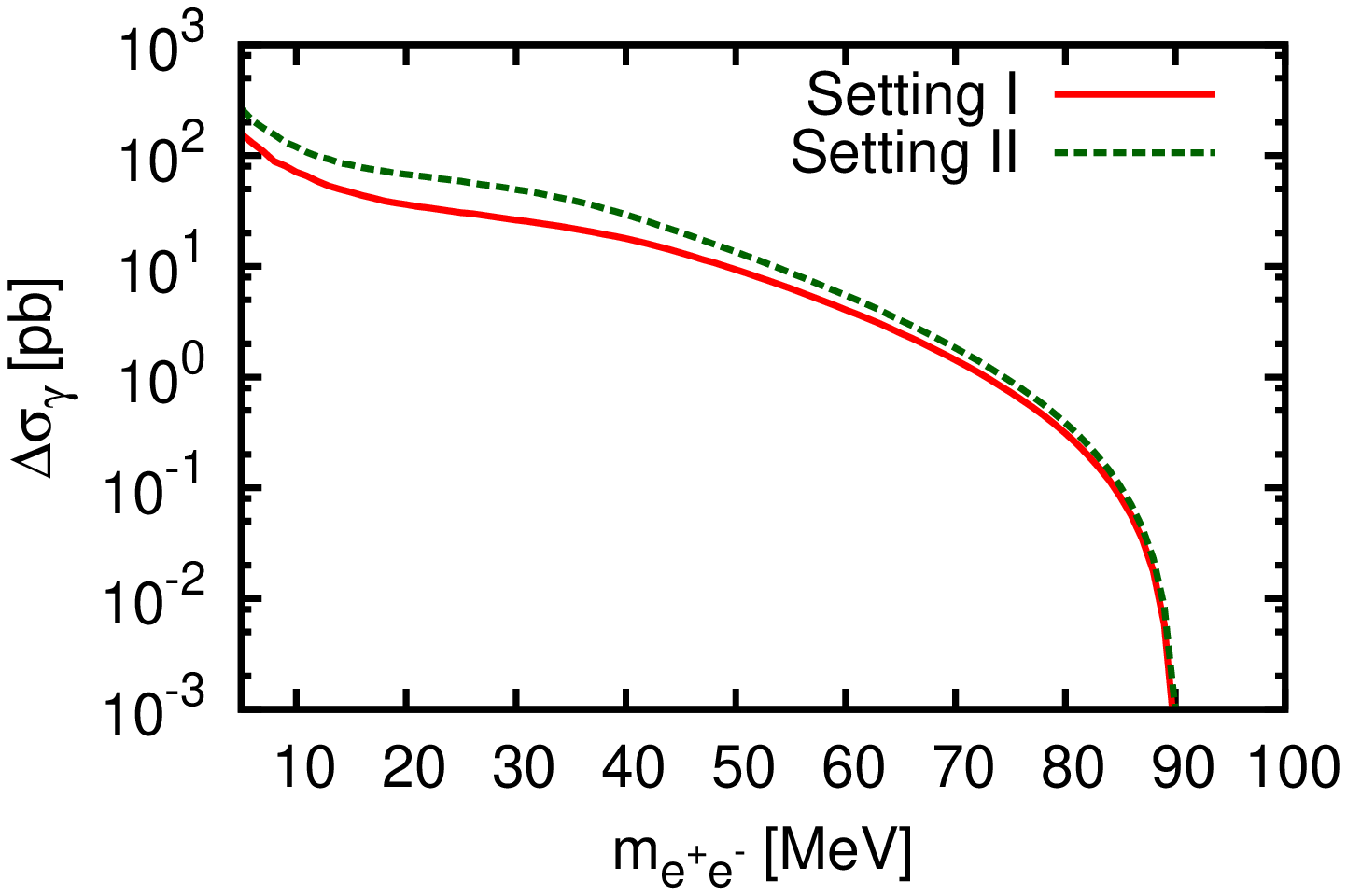} & \includegraphics[angle=-90,width=.49 \linewidth]{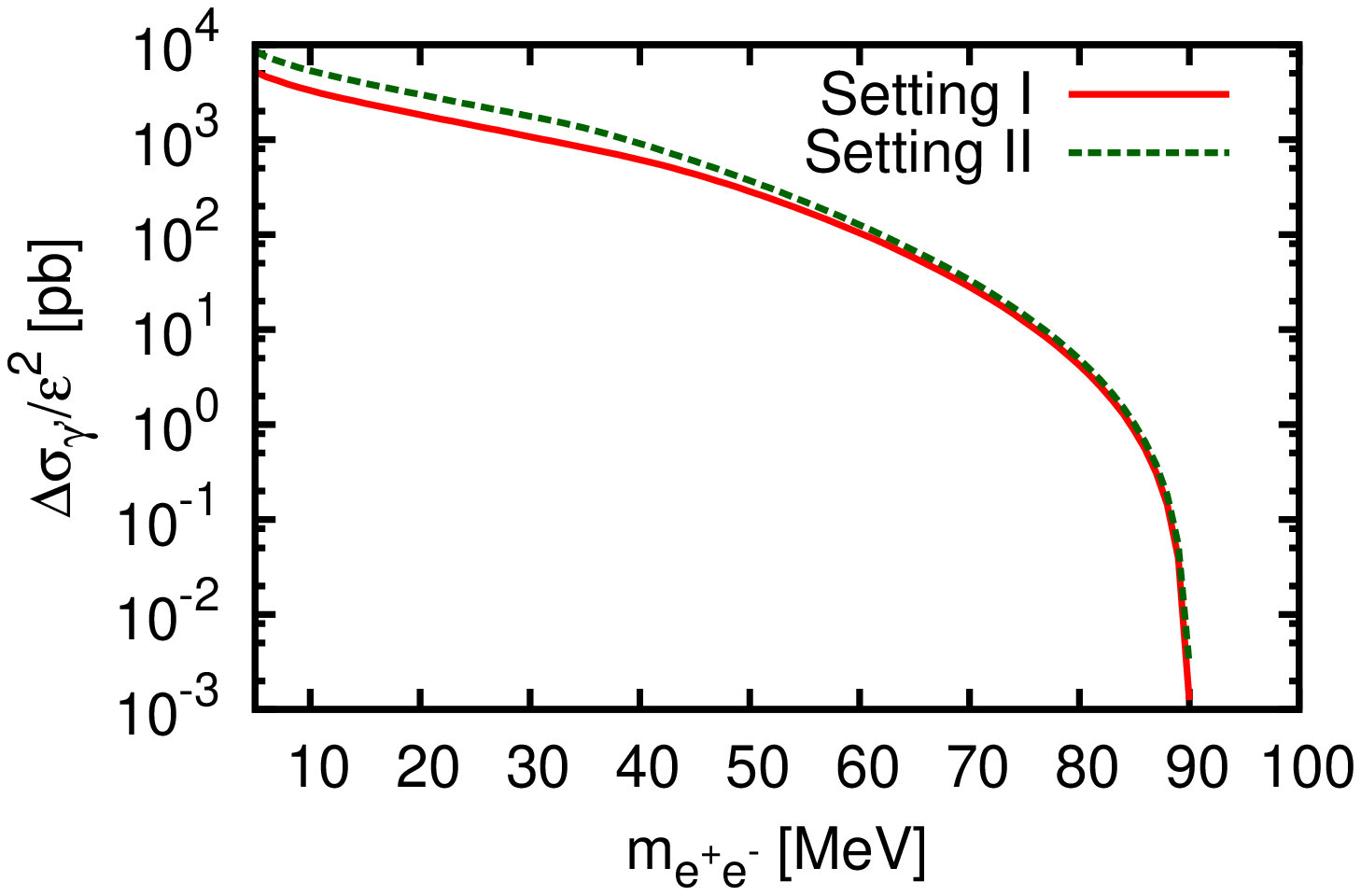} 
 \end{tabular}
\caption{Invariant mass distributions of the background (left) and signal normalized by $\varepsilon^2$ (right) cross section for DarkLight kinematics.
The solid (dashed) curve represents the calculation in Setting I (II) with values given in Tab.~\ref{tab:DarkLight_kin}.\label{fig:DarkLight_invmass}}
\end{figure*}
\begin{figure}[htb]
 \begin{tabular}{cc}
 \includegraphics[angle=-90,width=.49 \linewidth]{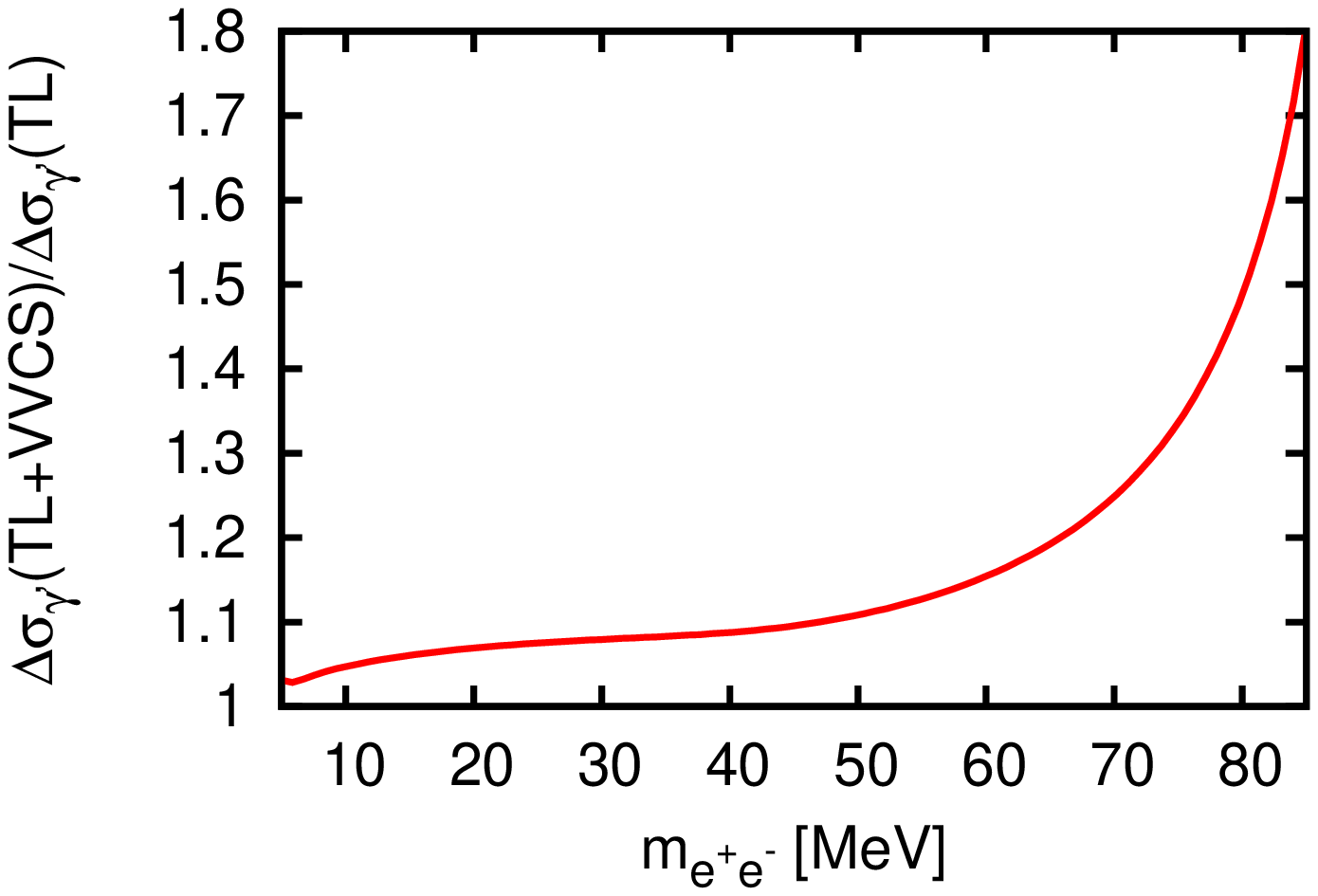}&
 \includegraphics[angle=-90,width=.49 \linewidth]{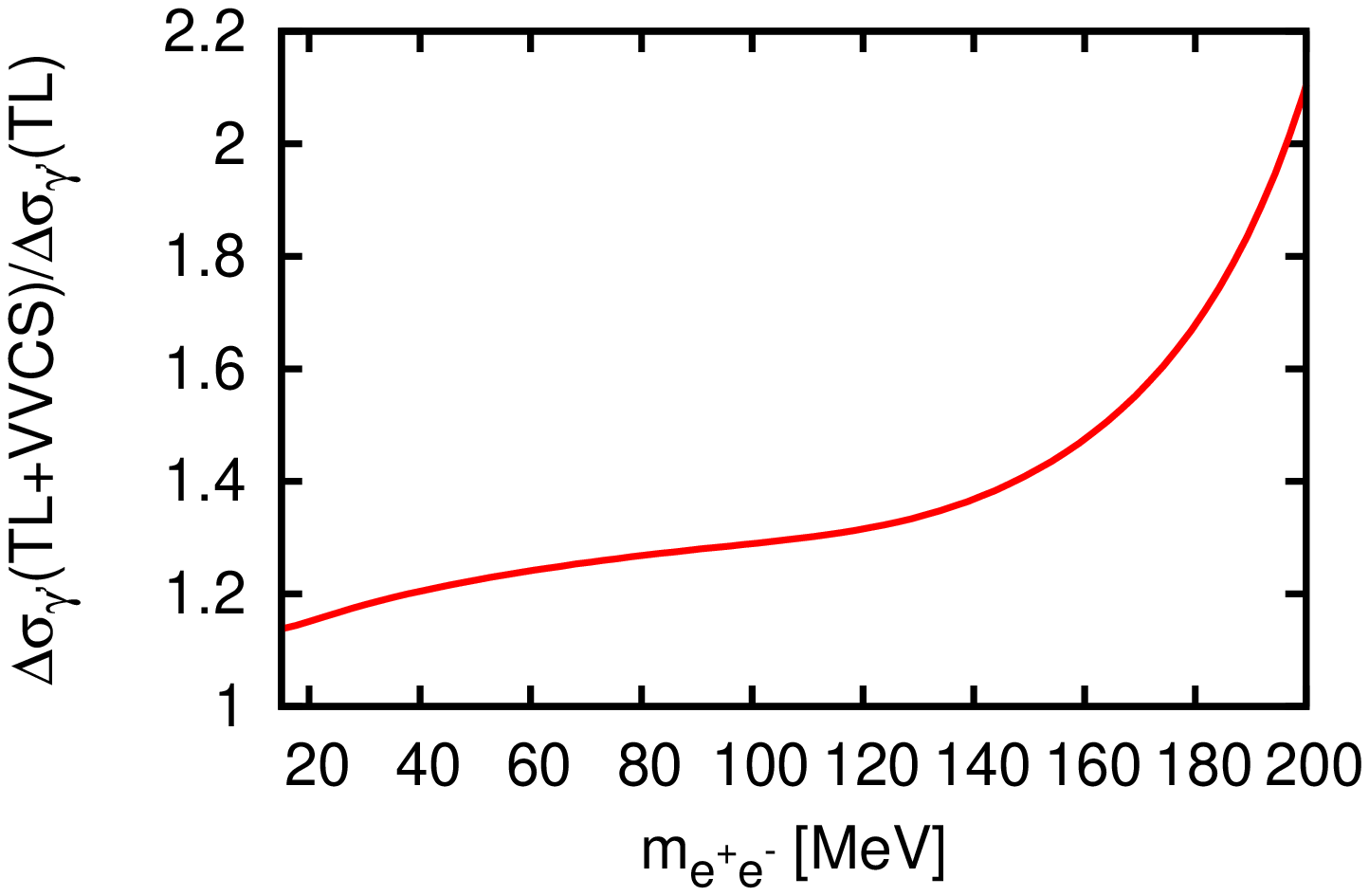}
 \end{tabular}
 \caption{Ratio of the signal cross sections for $\Ap$ production from the lepton, $\Delta \sigma_{\Ap}(TL)$, and additionally taking the VVCS contribution into account, $\Delta \sigma_{\Ap}(TL+VVCS)$, calculated in Setting I. Left panel: Beam energy $E_0=100\,\MeV$ as planned for the DarkLight experiment. Right panel: $E_0=300\,\MeV$.\label{fig:DarkLight_signal1vs2}}
\end{figure}

In the kinematical settings of Tab.~\ref{tab:DarkLight_kin} the cross sections necessary for the prediction of the reach of the experiment were calculated.
The results of the simulation of the QED background and the signal are shown in the left and right panel of Fig.~\ref{fig:DarkLight_signal1vs2}, respectively.
The distributions depending on the lepton pair invariant mass $m_{e^+e^-}$ using a bin width $\delta m = 1\,\MeV$ are shown.
As discussed in Sec.~\ref{sec:method}, for the production from a proton one has to take the VVCS contribution into account, which was done in this calculation for both, the background and signal cross sections.
Obviously, the cross section calculated in Setting II is slightly larger than the one in Setting I, which was expected due to the larger allowed phase space.
Note, that the signal cross section curve in the right panel was calculated assuming that the hidden photon mass is equal to the center mass of each bin.
This means, that the expected signal in the experiment is not the smooth distribution shown in the right panel of Fig.~\ref{fig:DarkLight_signal1vs2}, but only a signal in a single mass bin in which the hidden photon mass is contained.

Figure~\ref{fig:DarkLight_signal1vs2} shows the ratio of the signal cross section with $\Ap$ production from the lepton and proton ($\Delta \sigma_{\Ap}(TL+VVCS)$) and only from the lepton ($\Delta \sigma_{\Ap}(TL)$) for $E_0=100\,\MeV$ (left panel) and $E_0=300\,\MeV$ (right panel).
It turns out, that the VVCS contribution leads to an enhancement of the signal cross section for the full considered mass range, which is mostly in the range from 10 to 20\%.
For a larger hidden photon mass $m_\Ap$, the enhancement of the cross section grows and leads to a twice larger value.
Compared to studies of virtual Compton scattering off the proton \cite{Vanderhaeghen:2000ws}, where the process $ep\rightarrow ep\gamma$ is investigated, this enhancement appears unnaturally large.
This can be understood from the fact, that the cross section is suppressed by the mass of the radiated particle and the fermion mass in the propagator.
In the process $ep\rightarrow ep\gamma$ the bremsstrahlung radiation of a photon from the lepton is strongly peaking, while the radiation from the proton is suppressed by the comparatively large mass of the proton.
In the process $ep\rightarrow epe^+e^-$ the virtual (hidden) photon has a mass itself, which suppresses the peaked structures in the cross section.
As consequence, in particular for larger hidden photon mass, the VVCS contribution appears larger compared to the other contributions.
Furthermore, it is shown in the right panel of Fig.~\ref{fig:DarkLight_signal1vs2}, that the sharp upturn in the cross section ratio is caused by the particular choice of kinematics.
To illustrate this, we have calculated the cross section ratio for the same allowed angular acceptance but with a beam energy $E_0=300\,\MeV$.
One can clearly see, that for this beam energy the upturn in the ratio starts for much larger invariant mass of the lepton pair as in the case $E_0=100\,\MeV$.

\begin{figure}[htb]
 \begin{center}
  \includegraphics[angle=-90,width=.48\linewidth]{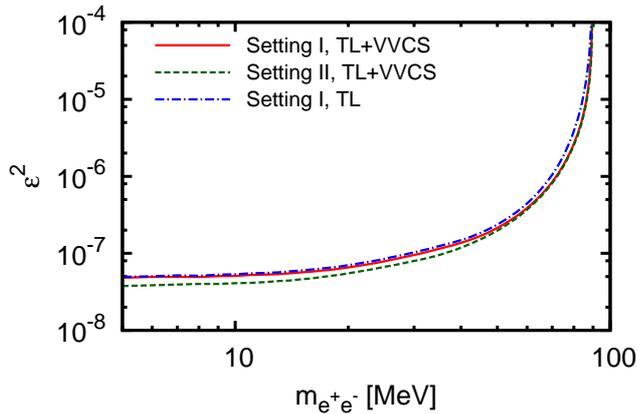}
 \end{center}
\caption{Projected reach of the DarkLight experiment assuming that a luminosity $L=1\,\mathrm{ab}^{-1}$, a mass resolution $\delta m = 1\,\MeV$ and a perfect detector efficiency can be reached.
The area above the curves can be excluded.
We show the predictions obtained in settings I (solid and dashed dotted curve) and II (dashed). 
In Setting I the exclusion limit was calculated using the TL + VVCS (solid) and the TL (dashed-dotted) signal cross section. In Setting II only the limit using the TL + VVCS signal cross section is presented.
\label{fig:DarkLight_limit_prediction}}
\end{figure}
Using Eq.~(\ref{eq:exclusionlimit}) and the calculated cross sections shown in Fig.~\ref{fig:DarkLight_invmass} we obtain predictions for the reach of the DarkLight experiment.
Following Ref.~\cite{Reece:2009un}, we assume that a perfect detector efficiency, a mass resolution $\delta m = 1\,\MeV$, and a luminosity $L=1\,\mathrm{ab}^{-1}$ can be reached.
Our results are presented in Fig.~\ref{fig:DarkLight_limit_prediction}.
Besides the prediction of the exclusion limit using the TL + VVCS signal cross section $\Delta \sigma_{\Ap}(TL+VVCS)$ illustrated by the solid (dashed) curve for Setting I (II), we show for comparison the exclusion limit derived from $\Delta \sigma_{\Ap}(TL)$ calculated in Setting I (dashed-dotted curve).
It turns out, that the exclusion limit is only less affected by the contribution of the VVCS process, whereas it has a significant effect on the cross section in the considered kinematical setting itself.
The small effect on the exclusion limit results can be explained by the smallness of the cross section itself where the VVCS effect becomes large.
Furthermore one can see, that the exclusion limit is also only slightly affected by the choice of the two kinematic settings.

We find, that our calculation leads to a similar projected reach as the one given in Ref.~\cite{Reece:2009un}, where the reach found in Ref.~\cite{Reece:2009un} is weaker by a constant factor of roughly 3.
This is due to the different choice of $N_\sigma$ in Eq.~(\ref{eq:exclusionlimit}).
While in Ref.~\cite{Reece:2009un} $N_\sigma = 5$ was used for the reach estimate, we have used $N_\sigma = 2$ in this work for reasons of comparability with Ref.~\cite{Beranek:2013yqa}.
Since $N_\sigma$ enters linearly in Eq.~(\ref{eq:exclusionlimit}), the two results obviously are in good agreement.
Due to the larger cross section of Setting II and the enhancement of the signal cross section by taking the VVCS contribution into account, the projected exclusion limit can be improved by around 10 to 30 \% in the considered mass range.
\subsection{Predictions for HPS-type experiments\label{sec:hps_type}}
In this section we study experiments of the same type as the future HPS experiment at JLAB \cite{HPS}.
Since we do not treat the more complicated actual setup of the HPS experiment in total, we refer to the setups as ``HPS-type''.
Like in the A1 and APEX experiments, HPS is designed to detect the created lepton and anti-lepton.
In contrast to A1 and APEX this is not done by two large spectrometers, but instead by a rather small detector array aligned directly at the beam which allows for the investigation of very small scattering angles at which the hidden photon signal is expected to be largest. 
The resulting cuts are for the horizontal scattering angles $\left|\Theta \right| \leq 50\,\mathrm{mrad}$ and for the vertical out-of-plane angles $-60\,\mathrm{mrad}\leq \alpha \leq -15\,\mathrm{mrad}$ and $+15\,\mathrm{mrad}\leq \alpha \leq +60\,\mathrm{mrad}$, respectively, which is given in the proposal of the HPS experiment \cite{HPS}.
Furthermore it is required, that the sum of the energies of the detected leptons exceeds $80\%$ of the beam energy, i.e. $(E_+ + E_-)/E_0> 0.8$ and that energy of each detected lepton is larger than $500\,\MeV$.
It is further demanded, that the two leptons may not be in the same half of the detector, i.e. the associated vertical scattering angles have opposite signs.

\begin{figure*}[htb]
 \begin{tabular}{ccc}
  \includegraphics[angle=-90,width=.32\linewidth]{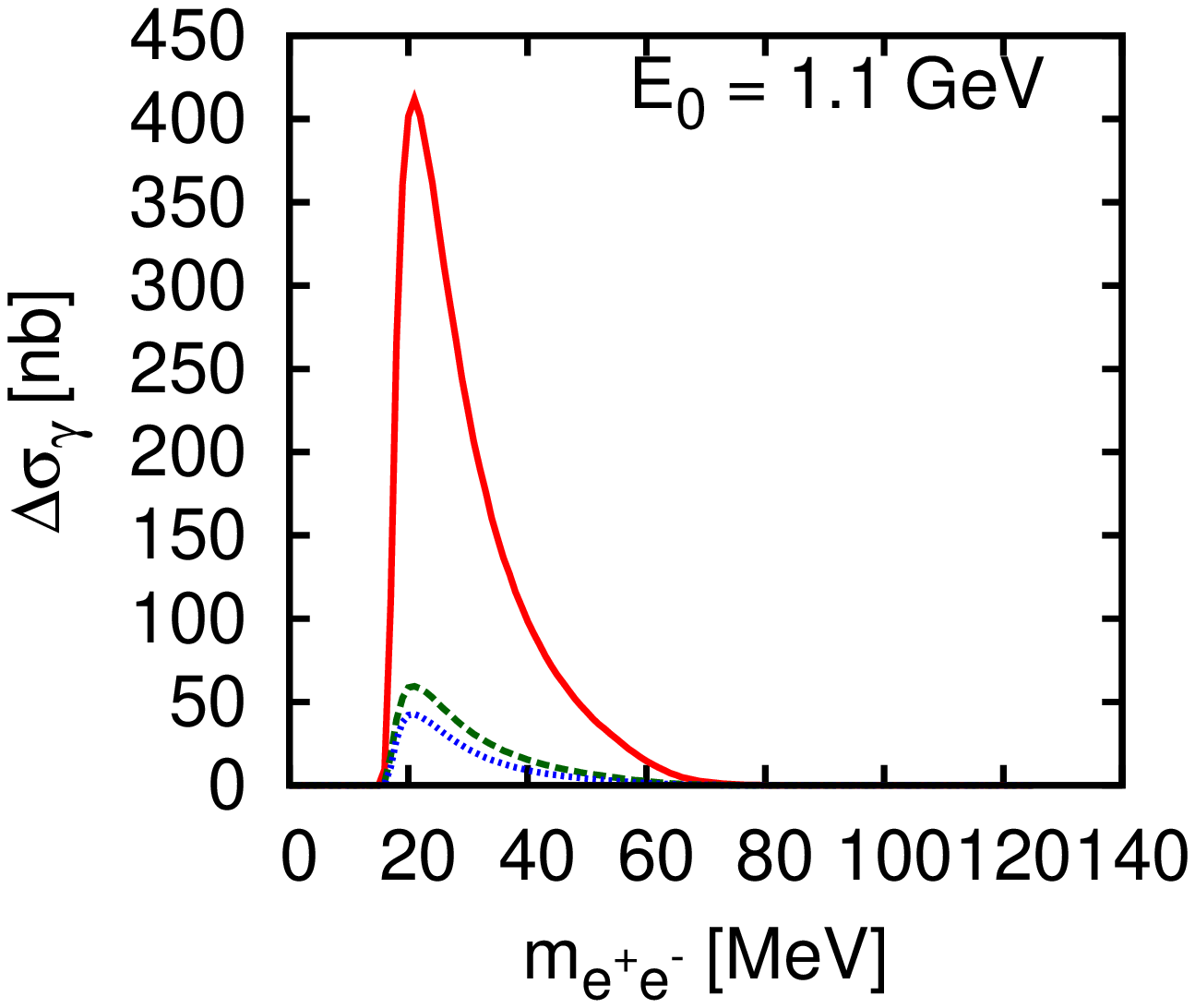}&
  \includegraphics[angle=-90,width=.32\linewidth]{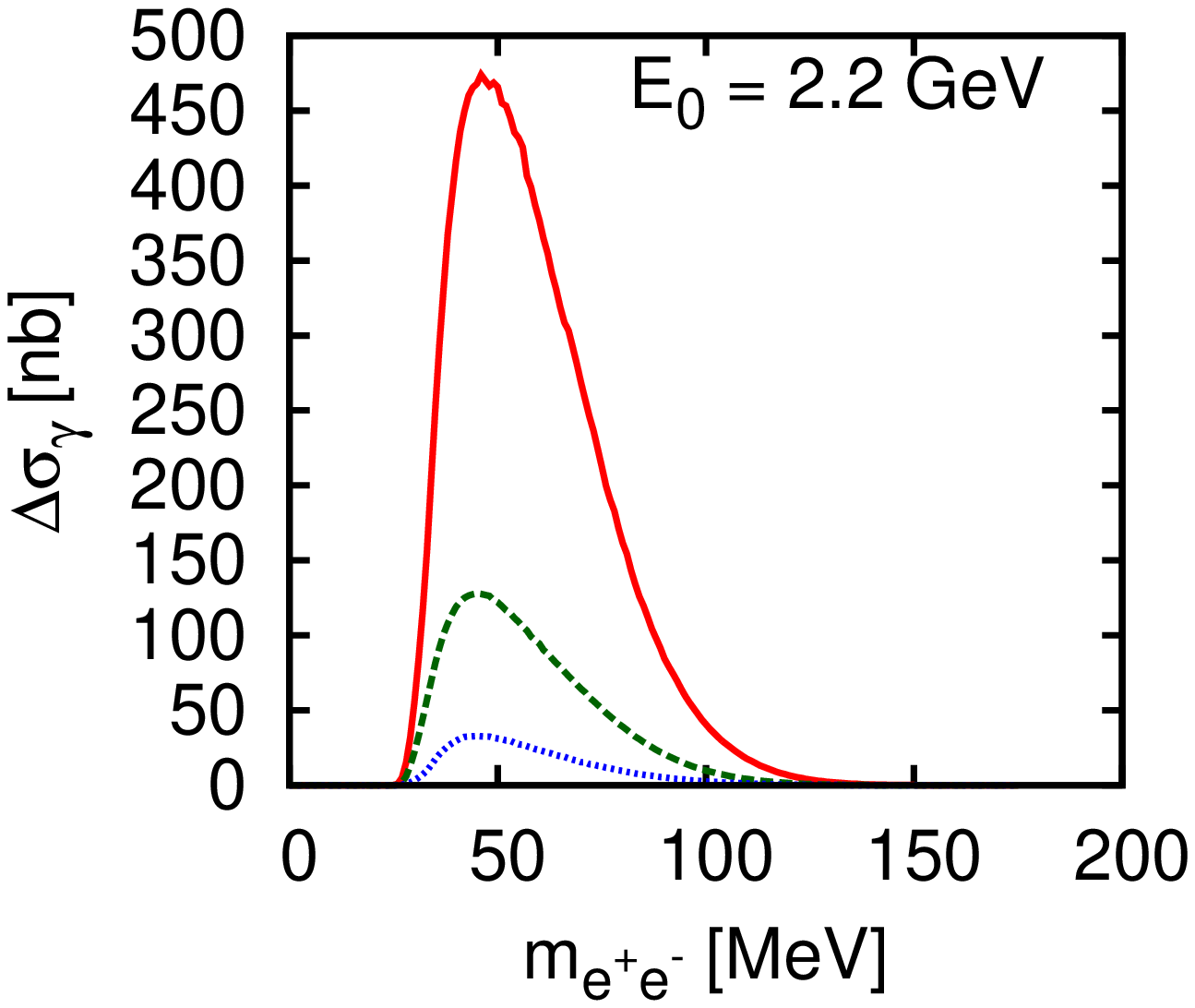}&
  \includegraphics[angle=-90,width=.32\linewidth]{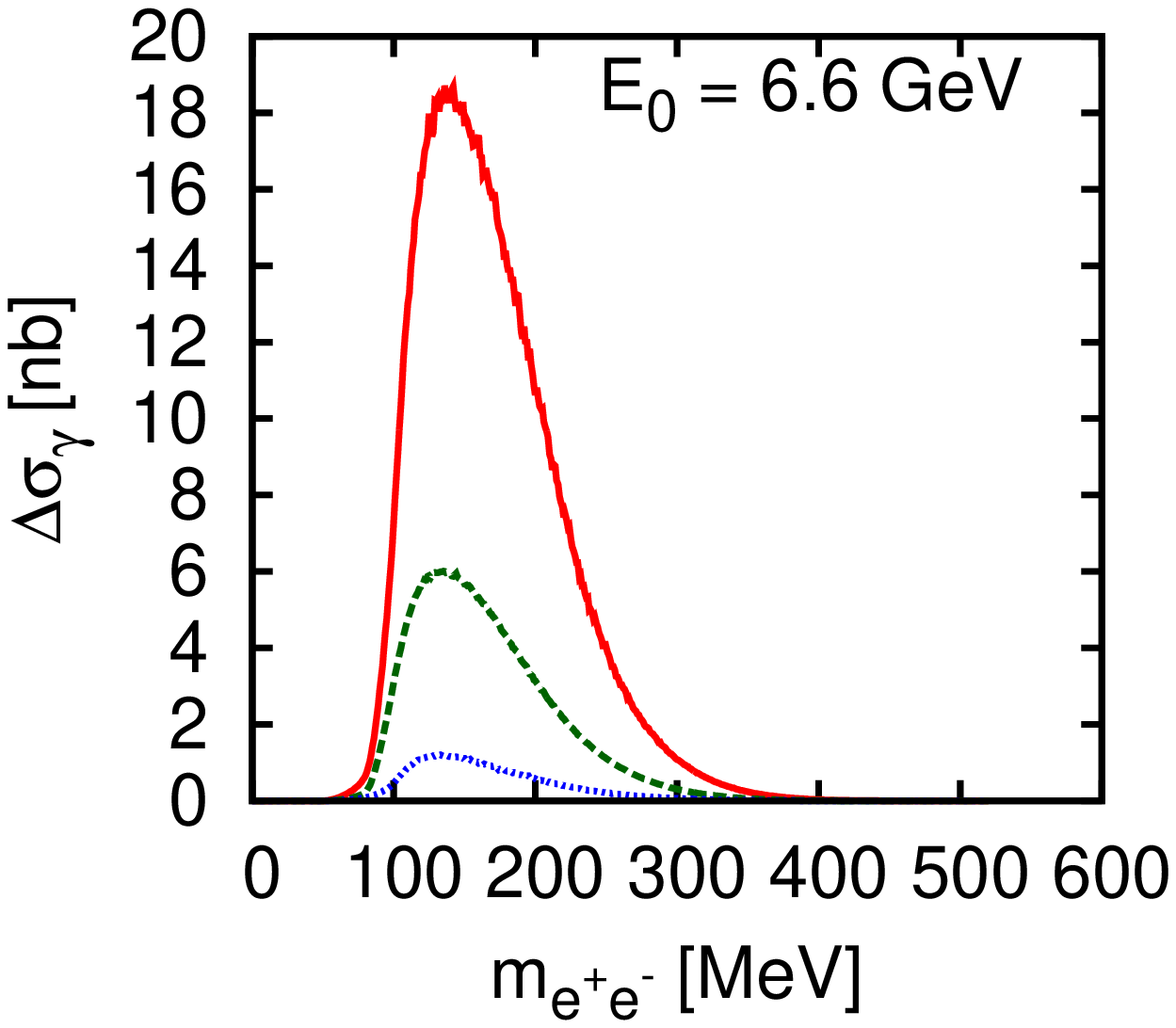}
 \end{tabular}
\caption{Invariant mass distributions of the background cross sections in HPS-type kinematics for selected beam energies. The TL+SL, D+X (TL+SL, D; TL, D) cross section is represented by the solid (dashed; dotted) curve.\label{fig:HPS_invmass}}
\end{figure*}
Following the proposal of the HPS collaboration \cite{HPS}, we calculate the cross section for beam energies $E_0=1.1\,\GeV$, $2.2\,\GeV$, and $6.6\,\GeV$.
The results of this simulation for HPS-type experiments can be found in Fig.~\ref{fig:HPS_invmass}, where the background cross section $\Delta \sigma_\gamma^{\mathrm{TL+SL,D+X}}$ ($\Delta \sigma_\gamma^{\mathrm{TL+SL,D}}$) is drawn by the solid (dashed) curve.
For comparison with other simulations we also show the the contribution to lepton-pair production from a timelike virtual photon $\Delta \sigma_\gamma^{\mathrm{TL,D}}$ (dotted curve), which is related to the signal cross section.
We find, that the largest part to the irreducible QED background originates from the contribution due to the indistinguishability of the final state electrons, which can be seen by comparing the curves for the cross sections $\Delta \sigma_\gamma^{\mathrm{TL+SL,D+X}}$ and $\Delta \sigma_\gamma^{\mathrm{TL+SL,D}}$.
Furthermore, the TL contribution compared to the SL one in the cross section with distinguishable final state electron becomes smaller for increasing beam energy, since the contribution from the intermediate timelike virtual photon is suppressed due to the increasing virtuality.
\begin{figure}[htb]
 \begin{center}
  \includegraphics[angle=-90,width=.48\linewidth]{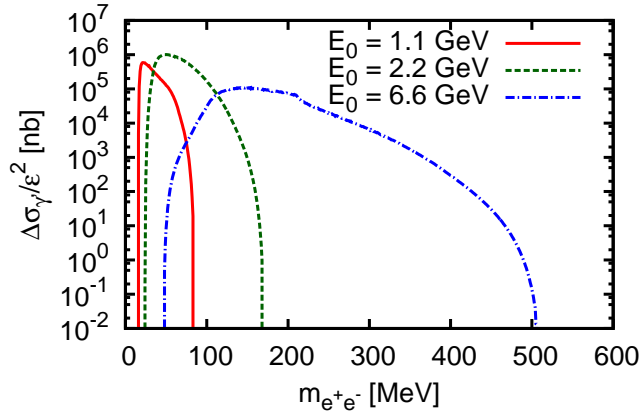}
  \caption{Signal cross section normalized by $\varepsilon^2$ in HPS-type kinematics.\label{fig:HPS_signal}}
 \end{center}
\end{figure}
The signal cross sections are shown in Fig.~\ref{fig:HPS_signal}.
Note, that this plot does not show the cross sections which will be observed in experiment but the theoretical curves for mass bins with $\delta m=1\,\MeV$, where the $\Ap$ mass is the center value of each bin.
Furthermore, the signal cross sections were divided by $\varepsilon^2$, which corresponds to setting $\varepsilon^2=1$.
The experimental signature of a $\Ap$ would be a peak at a certain mass with the height given in Fig.~\ref{fig:HPS_signal} over the smooth QED background presented in Fig.~\ref{fig:HPS_invmass}.

\begin{figure}[htb]
 \begin{center}
  \includegraphics[angle=-90,width=.48\linewidth]{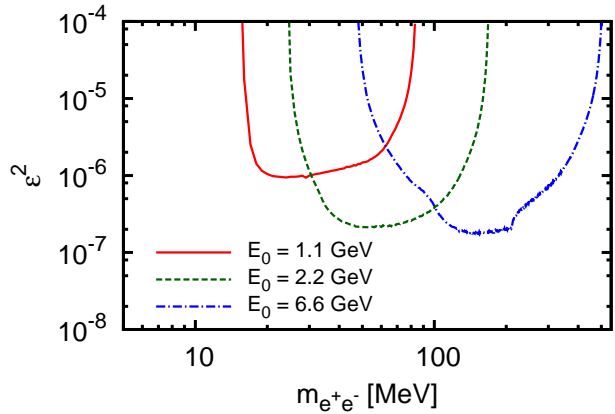}
 \end{center}
\caption{Projected exclusion limits for the investigated HPS-type kinematics. Integrated luminosities of $L= 5\,\mathrm{fb}^{-1}$, $L= 40\,\mathrm{fb}^{-1}$, and $L= 200\,\mathrm{fb}^{-1}$ were used for the settings with $E_0= 1.1\,\GeV$ (solid curve), $E_0= 2.2\,\GeV$ (dashed), and $E_0= 6.6\,\GeV$ (dashed-dotted), respectively, to estimate the reach.\label{fig:HPS_limit_prediction}}
\end{figure}
Again, we use Eq.~(\ref{eq:exclusionlimit}) and the obtained cross sections to find a prediction for the reach of a hypothetical experiment in the HPS-type kinematics.
For these calculations we assume a constant mass resolution $\delta m = 1\,\MeV$, which is better than the aimed resolution given in the HPS proposal, which depends on the beam energy and the invariant mass of the lepton pair.
Following the HPS proposal we insert integrated luminosities of  of $L= 5\,\mathrm{fb}^{-1}$, $L= 40\,\mathrm{fb}^{-1}$, and $L= 200\,\mathrm{fb}^{-1}$ for the settings with $E_0= 1.1\,\GeV$, $E_0= 2.2\,\GeV$, and $E_0= 6.6\,\GeV$, respectively, to estimate the reach of such an experiment.
Our results are presented in Fig.~\ref{fig:HPS_limit_prediction}, were the solid curve (dashed, dashed-dotted) shows the limit for a beam energy $E_0=1.1\,\GeV$ ($2.2\,\GeV$, $6.6\,\GeV$).
Note, that we show the individual bands only and do not give a prediction for the summed bins, which should increase the sensitivity significantly.
\subsection{Projection of exclusion limits}
\begin{figure}[htbp]
 \includegraphics[width=.75\linewidth]{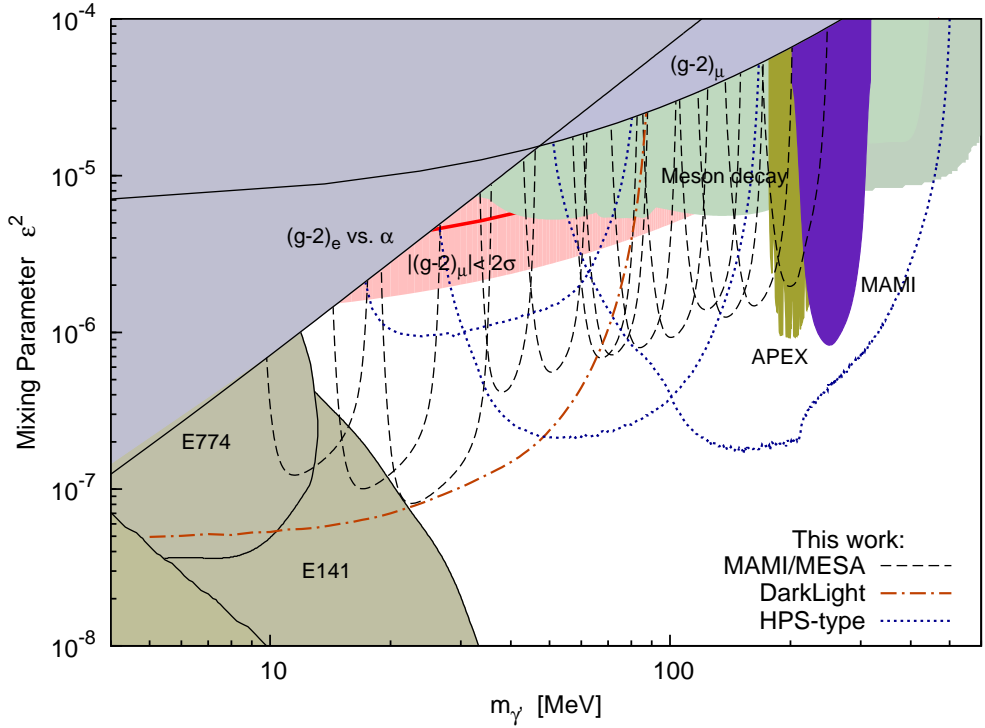}
 \caption{Summary of existing exclusion limits for visible $\Ap$ decays.
Existing limits as published in Refs.~\cite{Bjorken:2009mm,Davoudiasl:2012ig,Merkel:2011ze,Abrahamyan:2011gv,Endo:2012hp,Andreas:2011in,Archilli:2011zc,Babusci:2012cr,Gninenko:2013sr,Adlarson:2013eza} are shown, represented by the shaded regions.
The band denoted by $\left|(g-2)_\mu\right|<2\sigma$ is the $(g-2)_\mu$ welcome band, where the existing discrepancy between the experimental and theoretical value of the anomalous magnetic moment of the muon is most likely explained by the $\Ap$ contribution to $(g-2)_\mu$.
The results of our previous work~\cite{Beranek:2013yqa} for projected exclusion limits for MAMI and MESA are shown by the dashed curves.
The dotted  curves illustrate the projected bounds obtained in this work for an HPS-type experiment, where the bands correspond to the settings with beam energy $E_0=1.1\,\GeV$, $2.2\,\GeV$, and $6.6\,\GeV$ from left to right.
The prediction for DarkLight is shown by the dashed-dotted curve.
Note, that for a better visualization we do not show the predictions of other works \cite{Essig:2010xa,Freytsis:2009bh,HPS,Kahn:2012br}.\label{fig:exclusionplot_final}}
\end{figure}
Figure~\ref{fig:exclusionplot_final} shows a comparison of exclusion limits and projections for future bounds for the hidden photon parameter space for decays with purely visible decay products.
In this plot we show only the region of parameter space which is currently accessible at fixed-target experiments.
Existing limits are represented by the shaded regions \cite{Bjorken:2009mm,Davoudiasl:2012ig,Merkel:2011ze,Abrahamyan:2011gv,Endo:2012hp,Andreas:2011in,Archilli:2011zc,Babusci:2012cr,Gninenko:2013sr,Adlarson:2013eza}.
The $(g-2)_\mu$ welcome band denoted by $\left|(g-2)_\mu\right|<2\sigma$ corresponds to the region of parameter space, in which the current discrepancy between the theoretical and experimental determination of the anomalous moment of the muon can be explained by the contribution of the hidden photon.
The dotted and dashed-dotted curves in Fig.~\ref{fig:exclusionplot_final} illustrate the projected bounds obtained in this work for HPS-type experiments and the DarkLight experiment.
The individual bands for the HPS-type experiment correspond to three different settings with beam energy $E_0=1.1\,\GeV$, $2.2\,\GeV$, and $6.6\,\GeV$, respectively, from left to right.
For comparison, we also show the projected bounds obtained in our previous work~\cite{Beranek:2013yqa} for experiments at MAMI and MESA (dashed curve).
Note, that for a better visualization we do not show the predictions of other works \cite{Essig:2010xa,Freytsis:2009bh,HPS,Kahn:2012br}.
Furthermore, during finalizing this work, new limits from the HADES experiment \cite{Agakishiev:2013jla} excluding regions of parameter space in the mass range from $20$ to $60\, \MeV$ down to $\varepsilon^2 = 2.3\times 10^{-6}$ and from the $\nu$-calorimeter I experiment \cite{Blumlein:2013cua} were released, which are not yet included in Fig.~\ref{fig:exclusionplot_final}.

We find, that the experimental setups investigated in this work and the previous work will be conclusive whether or not a hidden photon decaying into a purely visible final state is at the origin of the discrepancy between the experimental and theoretical determination of the anomalous magnetic moment of the muon.
Furthermore, each of the investigated experiments will improve the existing current exclusion limit by at least one order of magnitude if no signal of a $\Ap$ can be found.
\section{Conclusion and Outlook\label{sec:outlook}}
In the first part of this work we have investigated the applicability of the Weizs\"acker-Williams approximation to calculate cross sections relevant for low-energy fixed target experiments searching for hidden photons.
We have found, that for beam energies above $5\,\GeV$ the shape of the cross section is well reproduced within the WW approximation.
For lower beam energies, the shapes differ significantly.
While from the WW approximation a steep rise of the cross section for $x\rightarrow 1$ is predicted, we find in our calculation, that the cross section sharply drops.
This result has a great impact for the actual design of experiments, since current experiments are designed such that the cross section for the emission of a hidden photon is largest, which is in the region $x\rightarrow 1$.
However, the formulae for the WW approximation found in Ref.~\cite{Bjorken:2009mm} overestimate the cross section calculated in leading order of QED by $30\%$ or more, also for large beam energies.
We have shown, that this is not an issue of the WW approximation itself but results from a too simplistic treatment of the hadronic current.
By a more sophisticated treatment, taking the angular dependence of the momentum transfer carried by the virtual photon into account, the calculation in leading order of QED and in the WW approximation agree within a few percent also for $x\rightarrow 1$, as long as the beam energy is sufficient large.
For beam energies, which are not much larger than the hidden photon mass, and $x\rightarrow 1$, both calculations still disagree.
However, one does not expect that the WW approximation is suited to describe the cross section in this region.

In the second part of this work we have applied the methods derived in our previous work~\cite{Beranek:2013yqa} to the kinematic settings of the DarkLight experiment and a setup of the type of the HPS experiment, which will both be performed at JLAB.
For that purpose we have calculated the signal and background cross sections within the experimental limits fixed by the detector geometry.
We have determined the region of parameter space which can be probed in the considered setups and have calculated the exclusion limits if no signal is seen.
We have found, that both experiments, such as the experiments proposed at Mainz, are ideally suited to probe the region of parameter space in which the $\Ap$ can explain the $(g-2)_\mu$ discrepancy.
Furthermore, these experiments can probe a region of parameter space, which extends more than one order of magnitude beyond the best current limits.
\begin{acknowledgments}
 This work was supported in part by the Research Centre ``Elementarkr\"afte und Mathematische Grundlagen'' at the Johannes Gutenberg University Mainz, the federal state of Rhineland-Palatinate, and in part by the Deutsche Forschungsgemeinschaft DFG through the Collaborative Research Center ``The Low-Energy Frontier of the Standard Model'' (SFB 1044) and the Cluster of Excellence ``Precision Physics, Fundamental Interactions and Structure of Matter'' (PRISMA).
 The authors thank R. Corliss, M. Guidal, M. Graham, H. Merkel, and P. Schuster for helpful discussions.
\end{acknowledgments}

\end{document}